\documentclass[aps,pre,singlecolumn]{revtex4}
\usepackage{graphicx}
\usepackage{mathptmx}
\usepackage{amsmath, amssymb}
\usepackage{bm} 
\usepackage[normalem]{ulem}
\usepackage{cancel}
\usepackage{nccmath} 

\usepackage{color}

\newcommand{\blue}[1]{{\color{blue} #1}}

\usepackage{amsfonts,amsmath,amssymb,amsthm}
\def\beq{\begin{equation}}
\def\eeq{\end{equation}}
\def\bea{\begin{eqnarray}}
\def\eea{\end{eqnarray}}
\usepackage[normalem]{ulem}
\usepackage{cancel}

\usepackage{hyperref}
\begin{document}
\makeatletter
\title{Dynamics and stability of inertial flexible chains under follower activity}

\author{Sattwik Sadhu}
\email{sat.sadhrsb@gmail.com}
\author{Nitin Kriplani}
\author{Raghunath Chelakkot}
\email{raghu@phy.iitb.ac.in}

\affiliation{Department of Physics,
Indian Institute of Technology
Bombay, Powai, Mumbai--400 076, India.}

\begin{abstract}{
\noindent

The dynamics of flexible polymers and chains under follower activity is known to produce diverse nonequilibrium states. A prominent feature of such systems is the emergence of periodic motion arising from the coupling between internal activity and chain conformation. Recently, it has been shown that flexible and extensible chains of active particles exhibit rich dynamical patterns in the overdamped limit, where inertia is negligible.
Here, we study the complex dynamics of a flexible and extensible chain of active particles under follower activity when inertia is significant. Using numerical simulations, we quantify the chain dynamics as a function of chain length ($N$), segment mass, and activity. To rationalize the numerical results, we develop theoretical descriptions in the limit of short chains ($N=3$) and long chains ($N \gg 1$). In both these limits, we derive approximate expressions for the bond lengths and bond angles along the contour, which show excellent agreement with the numerical results. In addition, for short chains, we derive the stability conditions for a periodic motion as a function of segment mass and activity.  For long chains ($N\gg1$) we identify parameter regime in which the circular, periodic solution becomes structurally unstable. Our theoretical and numerical analysis provides insights into the emergence of ordered and periodic behaviour in active chains.
}
\end{abstract}

\maketitle

\section{Introduction}
Collective dynamics of self-propelled and active particles give rise to a remarkable spectrum of dynamical patterns and emergent behaviors~\cite{attanasi2014finite, patel2022formation, kelley2013emergent,bialek2012statistical,cavagna2014bird,attanasi2015emergence, feliciani2016empirical, bacik2023lane, couzin2003self, murakami2019levy,franks2016social, bazazi2012vortex, delcourt2016collective,welch2001cell, allard2013traveling}. The variety of such patterns, found in synthetic as well as living systems, is often caused by the interplay between the activity of individual agents, inter-particle interactions, and multiple environmental factors~\cite{marchetti2013hydrodynamics, bechinger2016active}. Unlike their equilibrium counterparts, the interparticle interactions between active particles are not entirely derivable from a potential, as they are often modeled to mimic complex non-reciprocal interactions between the active entities. Some examples are, interaction between colloidal particles driven by phoretic forces, contact-induced interaction between cells, behaviour interaction between organisms, etc~\cite{lavergne2019group, gomez2022intermittent, duan2023dynamical, you2020nonreciprocity, dinelli2023non, fruchart2021non}. To study the collective behviour in such systems, theoretical and numerical studies have implemented effective interactions between active agents, which are a combination of reciprocal and non-reciprocal interactions~\cite{baconnier2025self, hiraiwa2020dynamic, das2024flocking}.    

One example of a non-reciprocal interaction implemented in numerical studies is the follower-force mechanism introduced in active chains and filaments
~\cite{chelakkot2014flagellar, de2017spontaneous, krishnamurthy2023emergent, chelakkot2021synchronized, elgeti2015physics, fily2020buckling, sangani2020elastohydrodynamical, Fatehiboroujeni2018, fatehiboroujeni2021three,thakur2022self, jain2022cargo, Laskar2013}. 
Similar to active-particle systems, active chains are driven by non-equilibrium forces that act locally along the chain. In the follower-force mechanism, the local active force is oriented at a fixed angle relative to the local tangent of the chain. Owing to this constraint, the direction of the local force evolves with the chain conformation. The resulting coupling between activity and chain geometry gives rise to a variety of periodic patterns, depending on the boundary conditions imposed on the chain and the strength of the active force relative to the chain stiffness. These patterns include flagella-like beating, spiralling, and rotational motion in two- and three-dimensional chains and filaments~\cite{chelakkot2014flagellar, elgeti2015physics, Fatehiboroujeni2018, fazelzadeh2023effects, anand2018structure, ghosh2014dynamics, isele2015self, peterson2020statistical, prathyusha2018dynamically,karan2024inertia,anand2019beating,kharayat2025kinetically, ishikawa2022pairing, ishikawa2025simple, tiwari2020periodic}

Recently, we analysed the patterns formed by an extensible and flexible chain of active particles in the overdamped limit, in absence of thermal fluctuations~\cite{sadhu2025active}. We showed that, in the overdamped limit, the chain exhibits a rich variety of periodic and quasi-periodic dynamics, depending on the chain length and activity strength. Remarkably, such periodic behaviour persists even in the shortest possible chain, consisting of only three segments. This recent study, however, indicated that the inclusion of inertia can significantly alter the chain dynamics, both qualitatively and quantitatively. In addition, several recent works in active and self-propelled systems have highlighted that introducing inertia can significantly alter the collective behaviour observed across several systems~\cite{lowen2020inertial, karan2024inertia, dai2020phase, sandoval2023free}.  Therefore, in the present work, we investigate the dynamics of follower-force-driven active chains with a particular focus on inertial effects. Furthermore, we develop a semi-analytical framework to describe the observed behaviour and demonstrate that its predictions are in good agreement with numerical simulations.

This article is structured as follows. Section II presents the model and its governing equations. In Section III, we discuss the results. We first derive general equations for rigid steady states(section III-A) and analyze the rigid-state solutions of a straight chain configuration (section III-B). Subsequently, we study the three-segment chain ($N=3$), derive analytical expressions for its dynamics, and compare the predictions with numerical simulations. We subsequently perform a stability analysis of the circular trajectories for $N=3$ (section III-C). In Section III-D, we derive approximate expressions describing the dynamics of long chains ($N\gg 1$) and compare the theoretical predictions with simulation results. Finally, Section IV summarizes the key findings and their implications.

\
\
\
\
\

\section{Model and methods}

\begin{figure}[h]
    \centering
\includegraphics[width=.5\textwidth]{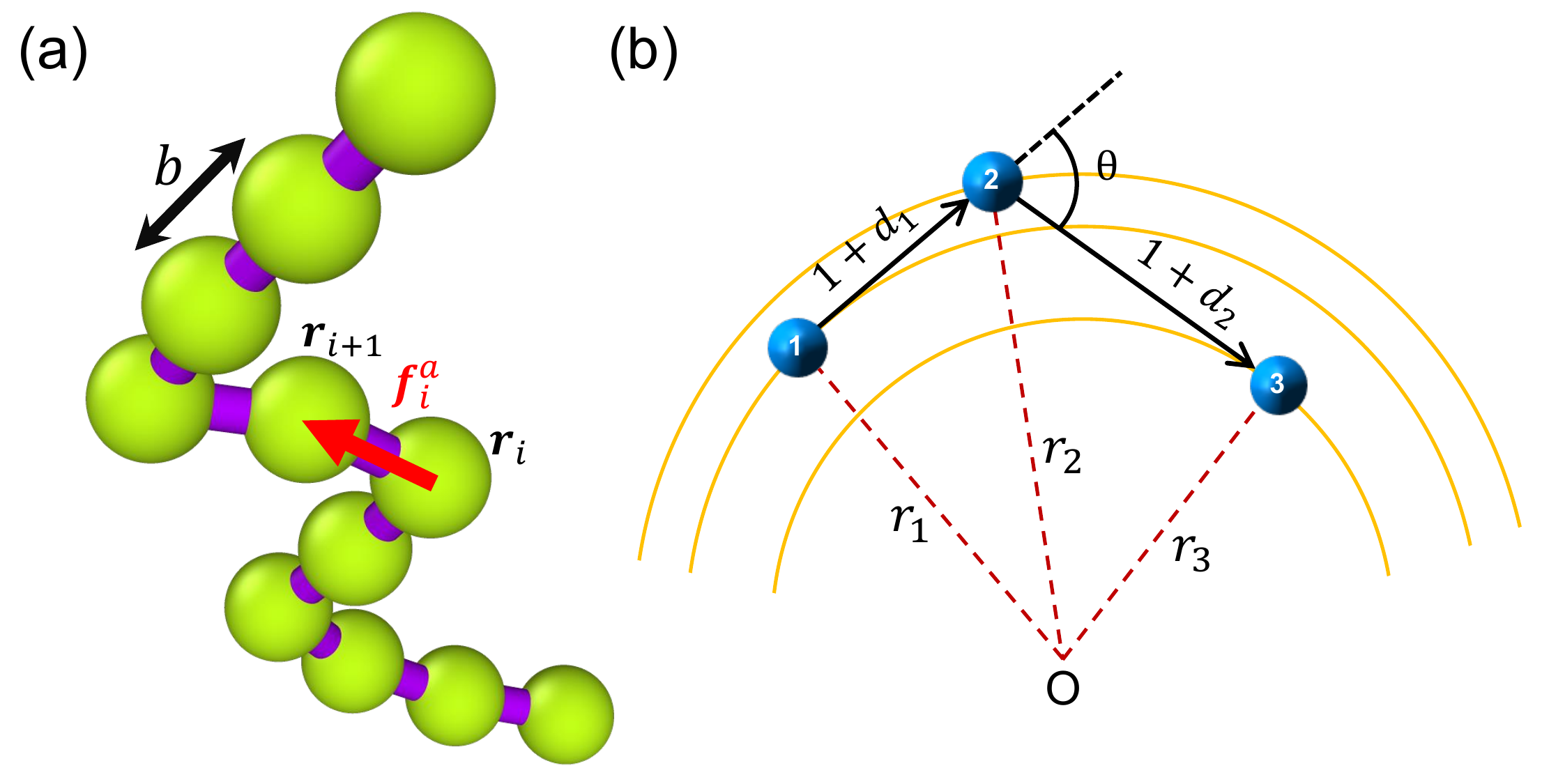}
    \caption{(a) Schematic showing the activity scheme used for our system under study (b) Schematic of the three bead system exhibiting circular steady state.}
    \label{fig1}
\end{figure}

The model for active chain contains $N$ particles (monomers), interacting via linear elastic springs, connected to form a flexible chain. A given monomer $i$ is linked to two of its neighbors via the potential, 
$V(\bm{r}_1,\dots,\bm{r}_N)=\frac{1}{2}k\sum_{i=1}^{N-1}(\lvert\bm{r}_{i+1}-\bm{r}_i\rvert-b)^2$, where $\bm{r}_i$ is the  $i^{th}$ particle position, $b$ the equilibrium bond length and $k$ the spring constant of the connecting bonds. Additionally, an active force $\bm {f}_i^a = f\hat{\bm{\Delta}}_i$ is acted on each monomer, along the unit tangent vector $\hat{\bm{\Delta}}_i$ at the $i^{th}$ monomer, which is defined as  $\hat{\bm{\Delta}}_i= (\bm{r}_{i+1} - \bm{r}_i)/|\bm{r}_{i+1} - \bm{r}_i|$, as shown in fig. \ref{fig1}(a). 

Thus, according to this scheme, each monomer acts as a polar active particle, whose self-propulsion direction is oriented along the local bond vector. 
This constraint couples the local activity of the chain to its conformation. Also, according to this definition, there is no active force applied on the leading monomer ($i=N$), and therefore, it remains passive.
In our previous study, we have analysed this system in the overdamped limit \cite{sadhu2025active}. Our study has revealed the existence of simple and complex dynamical steady states depending on the parameter values. Here we consider the full underdamped system with inertia, and the chain dynamics is governed by the equation $m\ddot{\bm{r}}_i(t)+\gamma\dot{\bm{r}}_i(t)=\bm{F}_i(t)+\bm{f}^a_i(t)$. The force term $\bm{F}_i=-\bm{\nabla}_iV$ is due to harmonic springs and $\bm{f}^a_i$ is the active, self-propulsion force as mentioned above. We have studied this system both analytically and numerically. For numerical simulations, we have used $b=m=1$ and $k=100\gamma$ unless specified otherwise.
 We used a second-order Velocity-Verlet scheme for simulation with timestep $\Delta t$ in the range $10^{-3}$ to $10^{-5}$, adjusted according to the parameter values to ensure consistency and convergence of the results. The numerical studies were conducted for values of $N$ ranging from $3$ to $1000$. We have started the simulations from different initial conditions, namely, random coil (in 2D and 3D) and several structured planar initial configurations, including zigzag and circular arrangement of segments. In the presence of inertia, we find that the steady-state configurations are independent of the initial conditions, irrespective of whether the system is initialized in two or three dimensions.
\

\section{results}
We have studied the chain dynamics by computing a range of parameters and tracking their temporal evolution. These parameters include, the bond-lengths $\Delta_i = {|{\bf r}_{i+1} - {\bf r}_i|}$,  bond angles $\theta_i=\cos ^{-1}(\bm{\hat\Delta}_i \cdot \bm{\hat\Delta}_{i+1})$, and the positions of the chain centre-of-mass (COM) and individual monomers. In the overdamped limit, we have observed the existence of two distinct types of steady-states. The first one is the rigid states, where both lengths $\Delta_i$ and the bond angles $\theta_i$ are constant in time. The second type of steady-state is flexible states, characterized by temporal variations in both these $\Delta_i$ and $\theta_i$. However, in the presence of inertia, the flexible steady-states disappear, and we predominantly observe rigid states for most of the parameter values. Therefore, in the subsequent sections, we develop the theoretical analysis for rigid states for chains in the presence of inertia. 
\subsection{General equations for rigid states}
A rigid steady-state is characterized by constant bond lengths $\Delta_i$ and bond angles $\theta_i$, while the overall chain orientation can vary with time. Also, the rigidity constraint implies that the chain orientation with respect to the net force on the chain is constant and the force also reorients with the chain. As a consequence, we obtain a circular trajectory when the chain reorients. The other possible state is a linear trajectory, when the orientation of the chain is fixed in space. As in the case of overdamped systems, the internal bond elastic force on each monomer $i$ can be written as $\bm{F}_i=k\{(\Delta_i-b)\hat{\bm{\Delta}}_i(1-\delta_{i,N})-(\Delta_{i-1}-b)\hat{\bm{\Delta}}_{i-1}(1-\delta_{i,1})\}$, where $\delta_{i,j}$ is the Kronecker delta function. 
To write the governing equations in a non-dimensional form, we scale the position and time such that, $\bm{r}_i\rightarrow b^{-1}\bm{r}_i$ and $t \rightarrow k\gamma^{-1}t$.
Subsequently, the non-dimensional forms for force $\tilde{f}=\frac{f}{kb}$ and mass $\tilde{m}=\frac{mk}{\gamma^2}$ are also introduced. Then, the governing equation reads,  
\begin{equation}\label{eq1}
    \tilde{m}\ddot{\bm{r}}_i(t)+\dot{\bm{r}}_i(t)=(d_i+\tilde{f})\hat{\bm{\Delta}}_i(1-\delta_{iN})-d_{i-1}\hat{\bm{\Delta}}_{i-1}(1-\delta_{i1}),
\end{equation}
where $d_i = \Delta_i -1$ is the bond extension of the $i^{\text{th}}$ bond.
To study the rigid states, we rewrite eq \eqref{eq1} in terms of the bond vectors ${\bm \Delta}_i$ providing
\begin{equation}\label{eq5}
m \ddot{\bm{\Delta}}_i + \dot{\bm{\Delta}}_i = (d_{i+1} + f ) \hat{\bm{\Delta}}_{i+1} - (2d_i + f ) \hat{\bm{\Delta}}_i + d_{i-1} \hat{\bm{\Delta}}_{i-1},
\end{equation}
for $i=1,\dots,N-1$ with $d_0=d_N+f=0$ (boundary conditions).
In the above equation, and in all the subsequent analysis, we drop the symbol $\tilde{(\cdot)}$ as all the parameters are in the dimensionless form. 
Since the rigid steady states are planar, we can assume that the motion happens in the X-Y plane without loss of generality, and thus express $\bm{\Delta}_i=\Delta_i(\cos \phi_i, \sin \phi_i)^T$, where $\phi_i$ is the angle between $\bm{\Delta_i}$ and the $x$ axis.
We also note that $\dot{\Delta}_i=0$ and $\dot{\phi}_i=\omega$ for $i=1$ to $N-1$ for rigid states by definition, where $\omega$ is the chain angular velocity. Taking the inner product with respect to $\hat{\bm{\Delta}}_i$ and its normalized derivative $\hat{\dot{\bm{\Delta}}}_i$ respectively on both sides of equation \eqref{eq5} , we get
%
%
%
\begin{subequations}\label{eq7}
\begin{align}
(d_{i+1}+f) \cos \theta_i-(2d_i+f-m\omega^2(d_i+1))+d_{i-1} \cos \theta_{i-1}  & = 0,\label{eq7a}\\
\text{and}\qquad\qquad\;(d_{i+1}+f) \sin \theta_i-(d_i+1)\omega-d_{i-1} \sin \theta_{i-1} & =0,\label{eq7b}
\end{align}
\end{subequations}
with $d_0=d_N+f=0$. 
We have used the definition of rigidity and bond angle to obtain \eqref{eq7}, which is valid for all the rigid states. 
Also, since we accept that $\dot{\Delta}_i =0$, the number of degrees of freedoms is lesser for these equations compared to eq.~\eqref{eq1}


\subsection{Solution for straight configuration}
As mentioned before, the orientation of the chain remains constant when the beads are initialized on a straight line. This is the trivial one dimensional unbound steady state, where $\theta_i$ can be either $0$ or $\pi$ for all $i$, thus giving
$\sin \theta_i=0$; $\cos \theta_i=\sigma_i=\{+1,-1\}$ and $\omega=0$, since the trajectory is linear. Hence we can solve the system \eqref{eq7a} to get the steady state properties. However, this steady state becomes mass independent once we put $\omega=0$ in \eqref{eq7a}, hence the solution is the same as obtained in the overdamped case \cite{sadhu2025active}, given by $d_i=-if/N$; $i=1$ to $N$ for the fully extended case ($\sigma_i=1$ for all $i$).
Mass in the system only determines how fast the system relaxes to this steady state. Summing both sides of eq. \eqref{eq1} over $i$ and dividing by $N$, we get $m\dot{v}_c+v_c=(1-\frac{1}{N})f$, where $v_c=\dot{r}_c$ is the velocity of the COM and $r_c$ is the position of the COM of the chain. If we assume that the chain started from rest, we obtain the solution $v_c(t)=v_\infty(1-e^{-t/m})$, where $v_\infty=(1-\frac{1}{N})f$ is the steady state velocity. Clearly, the system relaxes exponentially to the steady state with a dimensionless timescale $m$ which represents the amount of inertia in the system, and in the overdamped case, this relaxation is instantaneous. These observations are well known and expected and have been verified from our simulations as basic sanity checks.

\noindent
\subsection{Short chain with three segments}


\noindent
In this section, we specifically analyze the case of $N=3$, which is the simplest possible system that can be constructed within our framework(see Fig. \ref{fig1}(b)). We can explicitly write equations \eqref{eq7} for $i=1,2$ as follows
\begin{subequations}\label{eq8_1}
    \begin{align}
        (d_2+f)\cos\theta-(2d_1+f)+m\omega^2(d_1+1)=0,\label{eq8a_1}\\
        m\omega^2(d_2+1)-(2d_2+f)+d_1\cos\theta=0,\label{eq8b_1}\\
        (d_2+f)\sin\theta-(d_1+1)\omega=0,\label{eq8c_1}\\
        -(d_2+1)\omega-d_1\sin\theta=0,\label{eq8d_1}
    \end{align}
\end{subequations}
\noindent
where, we have replaced $\theta_1$ by $\theta$ and used the boundary condition $d_0=d_3+f=0$. Eliminating $\sin\theta$ and $\omega$ from eq~(\ref{eq8c_1}), and \eqref{eq8d_1}, we get
\begin{equation}\label{Eq10_1}
        d_1(d_1+1)+(d_2+f)(d_2+1)=0,
\end{equation}
which is exactly the same as what we obtained for the overdamped case \cite{sadhu2025active}. 
Adding and subtracting eq.\eqref{eq8b_1} from \eqref{eq8a_1} and \eqref{eq8d_1} from \eqref{eq8c_1}, we get
\begin{subequations}\label{eq9_1}
    \begin{align}
        m\omega^2(d_{+}+2)&=(d_{+}+f)(2-\cos\theta),\label{eq9a_1}\\
        (2-m\omega^2)d_{-}&=(f-d_{-})\cos\theta,\label{eq9b_1}\\
        (d_{+}+2)\omega&=(f-d_{-})\sin\theta,\label{eq9c_1}\\
        d_{-}\omega&=(d_{+}+f)\sin\theta,\label{eq9d_1}
    \end{align}
\end{subequations}
where $d_{\pm}=d_{1}\pm d_{2}$. Substituting $(d_{+}+f)$ from eq.\eqref{eq9d_1} in eq.\eqref{eq9a_1} and $(f-d_{-})$ from eq.\eqref{eq9c_1} in eq.\eqref{eq9b_1} and after doing some algebraic manipulation, we get 
\begin{equation*}
\frac{d_{-}}{d_{+}+2} = \frac{m\omega\sin\theta}{2-\cos\theta} = \frac{\omega\cos\theta}{\sin\theta(2-m\omega^2)},
\end{equation*}
or,
\begin{equation}\label{Eq10_2}
m \omega^2=2 - \frac{\cos \theta(2-\cos \theta)}{m \sin ^2 \theta}.
\end{equation}
\noindent
Using the value of $m\omega^2$ from eq.\eqref{Eq10_2} in \eqref{eq9a_1} and \eqref{eq9b_1} respectively, we get

\begin{subequations}\label{eq10_3}
    \begin{align}
        d_{+}=d_{1}+d_{2} &= \frac{(f-2)(2-\cos\theta)}{\cos\theta(1-\frac{2-\cos\theta}{m\sin^2\theta})} - 2 ,\label{eq10a_3}\\
        d_{-}=d_{1}-d_{2} &= \frac{f}{1+\frac{2-\cos\theta}{m\sin^2\theta}}.\label{eq10b_3}      
    \end{align}
\end{subequations}
\noindent
Thus, equations \eqref{Eq10_1}, \eqref{Eq10_2}, \eqref{eq10a_3}, and \eqref{eq10b_3} forms a closed system of equations, which in principle can be solved to get the bond lengths and bond angle for $N=3$. However, it is not possible to give an analytical closed form solution for these equations, hence we seek a good approximate solution, as explained further. To construct our approximation, we first look at the two limits, namely the overdamped ($m\to 0$) limit and strongly under-damped ($m\to\infty$) limit and then extend those solutions to approximate the behavior for the intermediate values of $m$.\\
\

\noindent
(i) Overdamped Limit: We know that the overdamped system ($m=0$) with $N=3$ has the solution $\cos\theta = 0$, $d_1 = d_2 = -\frac{f}{2}$ and $\omega = \frac{f}{2-f}$ \cite{sadhu2025active}. Hence, in this limit, we expect both $m\to0$ and $\cos\theta\to0$. This allows us to write $\cos\theta$ as a power series in $m$ as follows: $\cos\theta = a_1m + a_2m^2 + a_3m^3+\ldots$. and plug this expression in eq\eqref{Eq10_2} in the limit of $m\to0$. Since $\omega$ is finite in this limit, we obtain $a_1 = 1$ and $\omega^2\to1-2a_2$ as $m\to0$. Comparing this with $\omega = \frac{f}{2-f}$, we get $a_2 = \frac{2(1-f)}{(2-f)^2}$. Hence, in the overdamped limit, we have $\cos\theta = m+\frac{2(1-f)}{(2-f)^2}m^2+\mathcal{O}(m^3)$. However, using this limit argument, we cannot compute the higher-order coefficients ($a_3$ and above). This can be done by using a more general procedure as discussed in supplementary material (\hyperref[sec:sec3]{\blue{Supplementary S3}}). \\
\

\noindent
(ii) Strongly Underdamped Limit:
Let us assume the $\theta$ goes to a non-zero finite value (say $\theta^*$) as $m\to\infty$. Taking this limit in equations \eqref{eq10_3} and solving for $d_1$ and $d_2$, we get $d_1 = d_2 +f = \frac{(f-2)}{c^*}$, where $c^* = \cos\theta^*$. These values have to satisfy eq\eqref{Eq10_1}. Plugging in and solving, we get either $f=2$ or $c^*=2$. The former is impossible because $f=2$ implies $d_2=-2$ which in turn implies $\Delta_2 = -1$ which is unphysical as the bond length cannot be negative. The latter is impossible because the cosine of an angle cannot exceed $1$. 
Hence, by contradiction $\theta^*$ must be zero for $m \rightarrow \infty$. Thus, in this limit, we write $\sin^2\theta = \kappa m^{-1} + \mathcal{O}(m^{-2})$, which ensures that as $m\to\infty$, $\theta\to 0$ such that the term $m\sin^2\theta$ approaches a non-zero constant $\kappa$. Using this ansatz in equations \eqref{eq10_3} and taking the limit $m\to\infty$, we get $d_{1} = \frac{(f-1)\kappa^2 - \kappa}{\kappa^2-1} - 1$ and $d_{2} = \frac{(f-1)\kappa - \kappa^2}{\kappa^2-1} - 1$. Substituting these values in eq\eqref{Eq10_1}, we get the following condition on $\kappa$\\
\begin{equation}\label{eq10_4}
    (f-2)^2\kappa^3 + (f^2-6f+6)\kappa^2+f^2\kappa-(f^2-2f+2) = 0.
\end{equation}
This is a cubic equation which, in principle, can be solved to obtain the limiting value $\kappa$ for any given $f$. Although it is possible to obtain an exact, Cardano's solution for the above equation, it is too complicated for a physical interpretation. Moreover, this equation does not admit a single real solution for all values of $f$. Hence, we try to obtain a global approximate solution for $\kappa$.\\
\ 
\begin{figure*}[t]
    \centering
\includegraphics[width=0.85\textwidth]{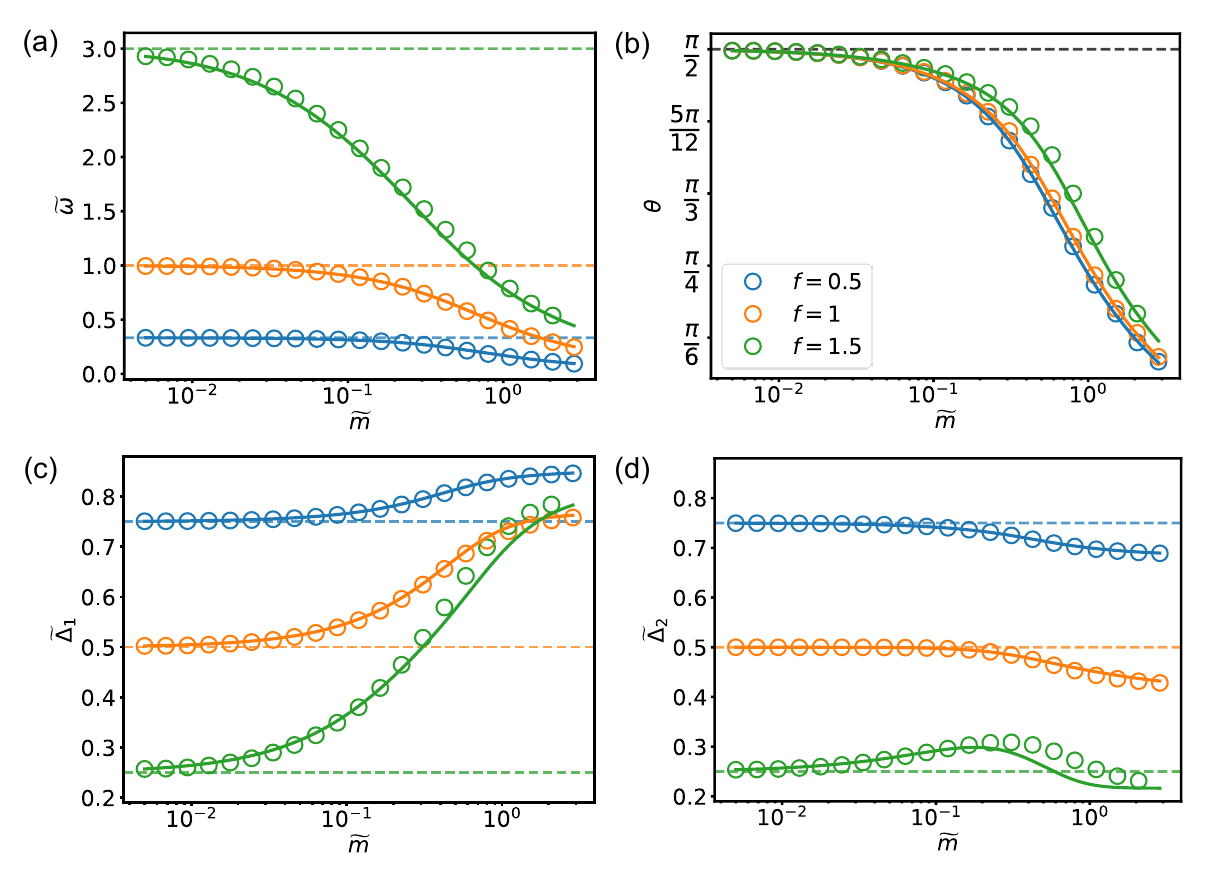} 
    \caption{Effect of inertia in the dynamics of a short chain with $N=3$. Variation of (a) Dimensionless angular velocity $\tilde\omega=\gamma k^{-1}\omega$, (b) Bond angle $\theta$, (c) Dimensionless bond length $\tilde\Delta_1=b^{-1}\Delta_1$ and (d) $\tilde\Delta_2=b^{-1}\Delta_2$ with the dimensionless mass $\tilde m=k\gamma^{-2}m$, where $\gamma$ is the damping constant, $k$ the spring stiffness and $b$ the equilibrium bond length. Blue, Orange and Green colors represent the dimensionless force $\tilde f=f/kb=0.5$, $1$ and $1.5$ respectively. The open circles represent the data points from simulation, the solid lines are the analytical solutions obtained using equation \eqref{padeww} in \eqref{Eq10_2}, \eqref{eq10_3} and $\kappa$ calculated using \eqref{pade}. The dashed lines correspond to the overdamped ($\tilde m=0$) solution which is already known \{$\tilde\omega=\tilde f/(2-\tilde f)$, $\theta=\pi/2$, $\tilde\Delta_1=\tilde\Delta_2=1-\tilde f/2$\}.}
    \label{fig2}
\end{figure*}
\noindent
It can be shown (see \hyperref[sec:sec1]{\blue{Supplementary S1}}), that a physically feasible solution exists only for $f\in [0, 2]$ and in this range $\kappa \in [\max(0.5, \frac{1}{3-f}), 1]$ with $\kappa (f=0) = 0.5$ and $\kappa (f=2) = 1$. Since we now have the bounds for the feasible roots, we can easily estimate $\kappa$ through any numerical method in this bounded domain (for example Newton-Raphson iterations starting from $k_{0} = 0.5$ converge pretty fast). Alternatively, we can also get simple closed-form approximate expressions in the form of Pad\'e approximants, etc (see \hyperref[sec:sec2]{\blue{Supplementary S2}}). The following Pad\'e $[3,4]$ approximation works very well with a maximum relative error of less than 0.1\% compared to the exact solution obtained numerically.
\begin{equation}\label{pade}
    \kappa^{pade}_{[3,4]}(f) = \frac{8(f^3-8f^2 + 18f - 18)}{f^4+16f^3-120f^2+288f-288}.
\end{equation}

Now, with the information of the two limits, we can proceed to construct a Pad\'e approximant for the variables of interest. We find that $\omega^2(m)$ can be used to express relevant system variables, such as bond lengths and bond angles. 
Hence, we will approximate $\omega^2(m)$ using a suitable Pad\'e approximant and then use that form to generate expressions for all other variables using equations \eqref{Eq10_2}, \eqref{eq10_3}.\\

To construct an approximate expression, we write the  Maclaurin (small $m$) expansion of $\omega^2$ as $\omega^2(m) = \sum_{l} a_{l}m^{l}$ and the asymptotic (large $m$) expansion be $\omega^2(m) = \sum_{l}b_{l}m^{-l}$, where $l\ge0$. The coefficients $a_l$, $b_l$ can be read off from equations (S13) of \hyperref[sec:sec3]{\blue{Supplementary S3}} as $a_0=F$, $a_1=-F(1+F)$, $a_2=F(5F^2+6F-3)/4$, $a_3=F(3+9F-9F^2-7F^3)/4$, $b_0=0$, $b_1=2-\kappa^{-1}$ and $b_2=\frac{1+(F-1)\kappa-(F+2)\kappa^2}{3(F+1)-4F\kappa+(F-3)\kappa^2}$, where $F = \frac{f^2}{(2-f)^2}$ is the square of the overdamped angular velocity. Using these coefficients, we can generate a $[2,3]$ Pad\'e approximant for $\omega^2$ using the general procedure described in \hyperref[sec:sec2a]{\blue{Supplementary S2.1}} as follows -
\begin{equation}\label{padeww}
    \omega^2_{Pade}=\frac{b_1pm^2+(a_0r+a_1)m+a_0}{pm^3+qm^2+rm+1},
\end{equation}
where 
\[p=-\frac{a_1^3 - 2 a_0 a_1 a_2 + a_0^2 a_3 - a_2^2 b_1 + a_1 a_3 b_1}{ a_0^3 + 2 a_0 a_1 b_1 + a_2 b_1^2 - a_1^2 b_2 + a_0 a_2 b_2},\]
\[q=\frac{a_0 a_1^2 - a_0^2 a_2 + a_1 a_2 b_1 - a_0 a_3 b_1 - a_2^2 b_2 + a_1 a_3 b_2}{ a_0^3 + 2 a_0 a_1 b_1 + a_2 b_1^2 - a_1^2 b_2 + a_0 a_2 b_2},\] and \[r=-\frac{a_0^2 a_1 + a_1^2 b_1 + a_0 a_2 b_1 + a_3 b_1^2 - a_1 a_2 b_2 + a_0 a_3 b_2}{ a_0^3 + 2 a_0 a_1 b_1 + a_2 b_1^2 - a_1^2 b_2 + a_0 a_2 b_2}.\] This expression looks cumbersome, but it turns out that a lower order Pad\'e (say, $[2,2]$) is insufficient to capture the observed non-monotonic behavior of $\Delta_2$ for large values of $f(\gtrsim1)$ as shown in Figure \ref{fig2}(d).\\

Using equation \eqref{padeww}, we can invert \eqref{Eq10_2} to obtain
\begin{equation*}
    \cos\theta=\frac{1-\sqrt{1-m(2-m\omega^2_{Pade})(1-m(2-m\omega^2_{Pade}))}}{1-m(2-m\omega^2_{Pade})},
\end{equation*}
taking only the feasible root, which can subsequently be used in equations \eqref{eq10_3} to obtain $d_1$, $d_2$. In Figure \ref{fig2}, we have compared the bond lengths $\Delta_1$, $\Delta_2$, bond angle $\theta$ and angular velocity $\omega$ between the simulation (open circles) and the analytical expressions following from \eqref{padeww} with $\kappa$ obtained from \eqref{pade} (solid line) as a function of $m$ for three different values of $f$, namely $f=0.5$, $1$ and $1.5$. It can be seen that the overdamped limit agrees with the expected values (dashed lines). The numerical simulations also provide good agreement in the underdamped limit as well ($m>0$). However, comparison in the highly underdamped limit ($m \gg 1$) is not provided since the simulations become unstable in that limit. This observation raises an interesting question about the stability of this system, which we explore in the following section.\\

Equations \eqref{eq7} do not give us information about the radii of the three beads straight away, but for our system, these can be expressed in terms of the known variables. To do so, we first note that by definition, a rigid steady state with fixed angular velocity $\omega$ implies that the position vectors $\bm{r}_i$ have a constant magnitude $r_i$, and without any loss of generality, we can assume the origin to be at the common centre of the concentric circular trajectories that the beads trace out, in which case $r_i$ simply becomes the radii of the beads. Also, $|\dot{\bm{r}}_i|=\omega r_i$, $|\ddot{\bm{r}}_i|=\omega^2 r_i$ and $\dot{\bm{r}}_i\cdot\ddot{\bm{r}}_i=\frac{d}{dt}(|\dot{\bm{r}}_i|^2/2)=0$. Moreover, by definition, $\cos\theta=\hat{\bm{\Delta}}_1\cdot\hat{\bm{\Delta}}_2$. Now, taking the square on both sides of equation \eqref{eq1} and simplifying, making use of these observations and the appropriate boundary conditions, we get for $i=1$, $2$ and $3$ respectively, $r_1=\frac{d_1+f}{\omega\sqrt{1+m^2\omega^2}}$, $r_2=\frac{\sqrt{d_1^2+(d_2+f)^2-2d_1(d_2+f)\cos\theta}}{\omega\sqrt{1+m^2\omega^2}}$ and $r_3=\frac{d_2}{\omega\sqrt{1+m^2\omega^2}}$. Hence, all system variables are now completely represented in terms of each other.

\begin{figure}[t]
    \centering
\includegraphics[width=.5\textwidth]{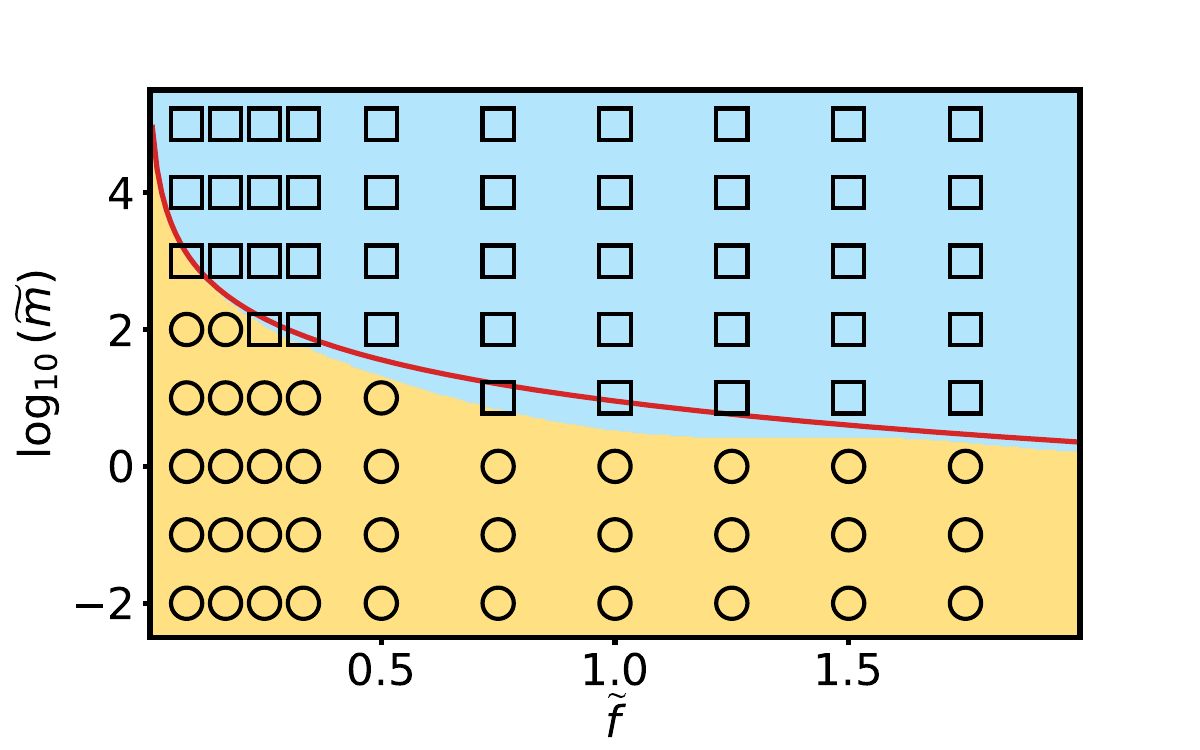}
    \caption{Stability Diagram showing the parameter region, where the 3-bead system is stable (amber) and unstable (sky blue). The x axis represents the dimensionless force $\tilde f=f/kb$, and the y axis represents the dimensionless mass $\tilde m=mk/\gamma^2$ in log scale. This diagram is constructed by numerically checking stability from eigenvalues of $J$ given by \eqref{jacobian}. The scatter points represents the steady state as obtained from simulations represented by open circles for circular steady states and open squares for non-circular trajectory. Simulations confirm that excepting a few `edge' cases, we indeed get a rigid circular state in the parameter regime where such a state is predicted to be stable. The solid red line represents the curve $\tilde m^*=9/\tilde f^2$, which is derived to be the approximate equation of the phase boundary for small $\tilde f$ and large $\tilde m$, but works surprisingly well throughout the entire feasible range of $\tilde f$}
    \label{fig3}
\end{figure}
\subsubsection{Stability of the rigid states}

For all the chosen values of $f$ in Fig.~\ref{fig2}, we were unable to find a rigid steady state from simulations for $m\gtrsim2.5$, although we know that there exists a rigid solution for all $m$ in the entire feasible range of $f$. This seems to suggest that such states for $m\gtrsim2.5$ might actually be unstable. Hence, it is crucial to find the physically relevant parameter regime, where the predicted rigid solution is actually stable. For the stability analysis, we go back to eq \eqref{eq5}, and again put $\bm{\Delta}_i=\Delta_i(\cos \phi_i, \sin \phi_i)^T$, and take the inner product with respect to $\hat{\bm{\Delta}}_i$ and $\hat{\dot{\bm{\Delta}}}_i$ as before, but this time without the rigidity assumption. We obtain,

\begin{equation}\label{eq9_2}
\begin{aligned}
    \dot{d}_{1} &= v_1, \quad \dot{d}_{2} = v_2, \quad \dot{\theta} = \omega_2-\omega_1,\\
    \dot{v}_1 &= (d_1+1)\omega_1^2+\frac{(d_2+f)\cos\theta-v_1-2d_1-f}{m},\\
    \dot{v}_2 &= (d_2+1)\omega_2^2+\frac{d_1\cos\theta-v_2-2d_2-f}{m},\\
    \dot{\omega}_1 &= \frac{(d_2+f)\sin\theta}{m(d_1+1)}-\left(\frac{2v_1}{d_1+1}+\frac{1}{m}\right)\omega_1,\\
    \dot{\omega}_2 &= -\frac{d_1\sin\theta}{m(d_2+1)}-\left(\frac{2v_2}{d_2+1}+\frac{1}{m}\right)\omega_2.
\end{aligned}
\end{equation}
\noindent
The Jacobian of this system \eqref{eq9_2} is a $7\times7$ matrix. By substituting the rigid steady state conditions ($v_1=v_2=0$, $\omega_1=\omega_2=\omega$) and simplifying via equations \eqref{eq8c_1} and \eqref{eq8d_1}, we obtain:

\begin{widetext}
\begin{equation}\label{jacobian}
    J = \begin{bmatrix}
        0 & 0 & 0 & 1 & 0 & 0 & 0 \\
        0 & 0 & 0 & 0 & 1 & 0 & 0 \\
        0 & 0 & 0 & 0 & 0 & -1 & 1 \\
        \frac{m\omega^2-2}{m} & \frac{\cos\theta}{m} & -\frac{(d_1+1)\omega}{m} & -\frac{1}{m} & 0 & 2(d_1+1)\omega & 0 \\
        \frac{\cos\theta}{m} & \frac{m\omega^2-2}{m} & \frac{(d_2+1)\omega}{m} & 0 & -\frac{1}{m} & 0 & 2(d_2+1)\omega \\
        -\frac{\omega}{(d_1+1)m} & \frac{\sin\theta}{(d_1+1)m} & \frac{(d_2+f)\cos\theta}{(d_1+1)m} & -\frac{2\omega}{d_1+1} & 0 & -\frac{1}{m} & 0 \\
        -\frac{\sin\theta}{(d_2+1)m} & -\frac{\omega}{(d_2+1)m} & -\frac{d_1\cos\theta}{(d_2+1)m} & 0 & -\frac{2\omega}{d_2+1} & 0 & -\frac{1}{m}
    \end{bmatrix}.
\end{equation}
\end{widetext}

\noindent
The Jacobian eq\eqref{jacobian} is a generic form which can be used for the stability analysis of both circular and linear steady states. Since $\omega =0$ for linear states, the Jacobian (eq\eqref{jacobian}) becomes  sparse so that the corresponding characteristic polynomial factorizes into smaller polynomials. Thus, it can be trivially shown that linear states are unstable in the domain of interest (see \hyperref[sec:sec4a]{\blue{Supplementary S4.1}}).\\

This leaves us with the circular state, whose stability we need to analyze. Simulations already indicate that the circular state is not stable for all parameter values. More precisely, the circular state is stable in the overdamped limit as expected (see \cite{sadhu2025active}), but becomes unstable as the value of $m$ is larger than a certain threshold (see figure \ref{fig3}). Ideally, we would like to find this critical mass, $m^*$ as a function of $f$. However, due to the high degree of nonlinearity in the system, it is difficult to analyze the system \eqref{eq9_2} with full generality. Hence, as before, we analyze the stability only in certain limits to deduce the approximate behavior for the entire parameter space and subsequently compare it with simulations. 

More specifically, we focus on the regime where $m$ is large and $f$ is small. In this parameter regime, the numerical analysis (Fig~\ref{fig3}) shows that the stability of circular states is relatively more sensitive to $f$ since the stability boundary shows a sharp decay in this parameter regime. The numerical results indicate that this decay is approximately of the form $m^* \propto f^{-\alpha}$, $\alpha>0$. To estimate the approximate value of $\alpha$, we analyze the stability of the system by constructing a Routh table from the characteristic polynomial of the Jacobian (eq.~\eqref{jacobian}). To analyze the stability of the system we employ the well-known Routh-Hurwitz criteria~\cite{routh,hurwitz}. We first construct a Routh table from the characteristic polynomial of the Jacobian \eqref{jacobian}. According to the Routh–Hurwitz criterion, a necessary condition for stability is that all coefficients of the characteristic polynomial, which appear in the first two rows of the Routh table, must have the same sign. The system is stable iff in addition to the necessary condition, it also satisfies the following sufficient condition that all elements in the first column of the Routh table have the same sign, i.e. there are no sign changes in the first column. (see \hyperref[sec:sec4b]{\blue{Supplementary S4.2}} for details) 

All the entries of the Routh Table are polynomials in $m$, $f$ and $\kappa$ of the form $m^pf^q\kappa^r$. $\kappa$ naturally comes because we are looking at the large $m$ (asymptotic) expansion of different variables. However, $\kappa$ has a dependence on $f$ of the form $\kappa=\frac{1}{2}+\frac{f^2}{72}+\mathcal{O}(f^3)$, in the small $f$ limit. Further, if we scale any particular row of the Routh Table in such a way that the leading term of the first entry of that row is $1$, then the first two dominant terms of the first entry of that row will be of the form $1+\nu m^pf^q$. If this $\nu$ is positive, the dominant term never changes sign, but if $\nu$ happens to be negative, there will be a change of sign provided that the second term is $\mathcal{O}(1)$ (which requires $p<0,\ q>0$). This condition naturally gives us the power law behavior, as hypothesized. Interestingly, doing this entire analysis (\hyperref[sec:sec4b]{\blue{Supplementary S4.2}}), we find that the sufficient condition for stability is indeed violated in one of the rows and $\nu$ is indeed negative in the corresponding row. Subsequently, we find that the phase boundary approximately behaves as $m^*=9/f^2$ for small $f$. The circular steady state is stable for $m<m^*$, and unstable otherwise. Although this approximation is derived for small $f$, it works well even for larger values of $f$, as can be seen from figure \ref{fig3}.

\subsection{Long chains with $N\gg1$}

\begin{figure*}[t]
    \centering
    \includegraphics[width=\textwidth]{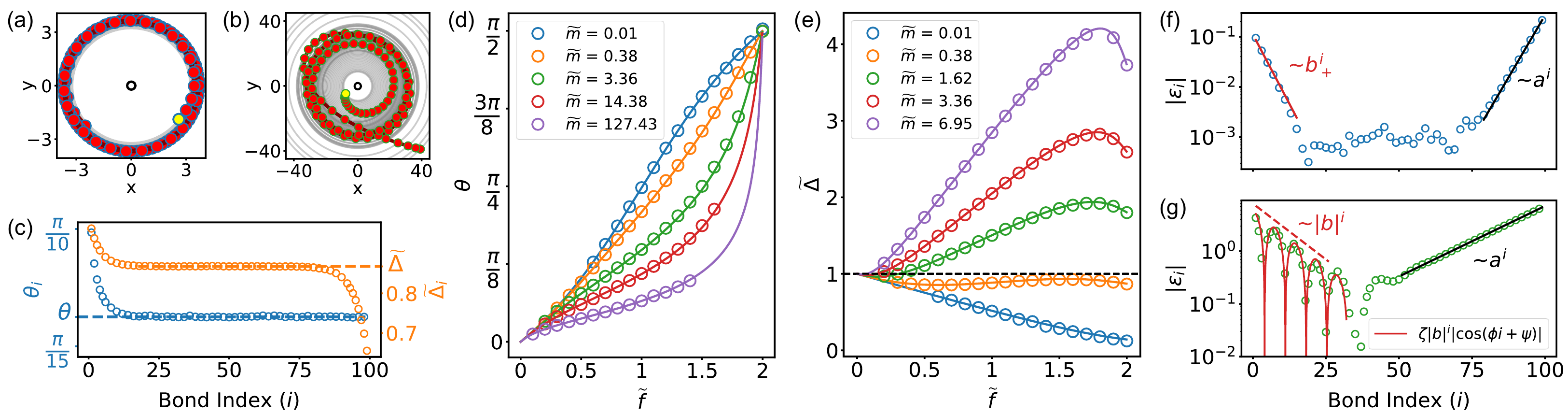}
    \caption{(a-b) Steady state chain configuration and trajectory for the $(\tilde m,\tilde f)$ pairs $(0.38,0.4)$ and $(61.6,0.9)$ respectively. $\tilde m=mk/\gamma^2$ is the dimensionless mass and $\tilde f=f/kb$ is the dimensionless active force. The light gray lines represent the individual particle trajectories and the black line represents the COM trajectory. (c) Dimensionless bond length ($\tilde\Delta_i$) and bond angle ($\theta_i$) profile for the configuration (a). The plateau regions correspond to the unperturbed stationary values $\tilde\Delta$ and $\theta$ respectively. (d-e) Plot of stationary values $\theta$ and $\tilde\Delta$ respectively (obtained as shown in (c)) with $\tilde f$ for different values of $\tilde m$ (open circles) and comparison with the theory (solid lines) obtained using the expressions \eqref{ii}-\eqref{iv}. (f-g) Profile plot of magnitude of bond length deviations ($\varepsilon_i$) in $\log$ scale for the configurations (a-b) respectively. Open circles represent data points from simulation and the solid lines represent theoretical predictions. Quantities $a$, $b_\pm$ and $\phi$ are computed from theory, while the constants $\zeta$ and $\psi$ in (g) are fitted from the data points.}
    \label{fig4}
\end{figure*}

Previous studies conducted by us in the overdamped limit ($m \rightarrow 0$) for chains with $N>3$ revealed a variety of bound and unbound, flexible dynamical states where the chain moves along non-circular trajectories~\cite{sadhu2025active}. These trajectories include complex, wave-like and spirographic patterns. 
However, when the inertial effects are significant ($m>0$), the flexible steady-states disappear and the chain typically form a circular configuration. For $N \gg1$ these configurations lead to self-intersecting loops, as the chain length is larger than the loop size (Fig~\ref{fig4}(a)-(b)). Interestingly, we observe two types of self-intersecting configurations depending on the parameter values while the chain follows a circular trajectory. In the first case (Fig~\ref{fig4}(a)), all the monomers are arranged approximately on a circular loop with nearly uniform curvature. In the second case, (Fig~\ref{fig4}(b)) the curvature varies significantly along its contour, with the tail end remaining almost straight. We show that this difference in the steady-state configuration is a manifestation of a subtle difference in the strain distribution along the chain.  

The presence of an excluded volume interaction between the segments avoids this self-intersection. But such interactions act as a source of interaction-induced noise, and the chain collapses into a random coil with a finite decay length, as observed in the previous studies~\cite{bianco2018globulelike, malgaretti2025coil}. Since the objective of this study is to understand the dynamical states in noiseless chains, we do not add any excluded volume interactions and hence the chains are allowed to intersect. Although these states are unrealistic, such an analysis reveals several interesting aspects about emergent periodicity and internal dynamics of chains during this periodic motion. We also note that, when the $m$ is very large, these circular steady states also become unstable, leading to a variety of unsteady and flexible states. However, we focus only in the parameter regime where such circular states are stable. 


From the simulations, we observed that the bond lengths and bond angles of the chain remain constant for most of the chain length with significant deviation only towards the ends as observed in Fig \ref{fig4}(c). Hence, it is reasonable to consider an index independent bond length $\Delta$ and bond angle $\theta$ which applies to those regions. Chain segments which deviate from this behaviour can subsequently be treated as perturbation to these constant values. To obtain these constants, we plug in $\Delta_{i} = \Delta$ and $\theta_{i} = \theta$ for all $i$ in equation \eqref{eq7}. Which yields the following relations\\

\begin{subequations}\label{relation16}
\begin{align}
(2d+f)(\cos\theta - 1) + m\omega^2(d+1)&=0, \\
\omega(d+1) = f &\sin\theta,
\end{align}
\end{subequations}
where $d = \Delta - 1$ is the bond extension.\\  
Now, to capture the deviation in bond length, we introduce the variable $\varepsilon_i=\Delta-\Delta_i=d-d_i$. At this point, we still assume $\theta$ to be constant, which is not an unrealistic assumption, because in Fig \ref{fig4}(c), we can see that towards the head end ($i\to100$), the bond lengths deviate significantly from $\Delta$ while still keeping the bond angles unperturbed. Using eq.\eqref{eq7} and relations \eqref{relation16}, we get
\begin{equation}\label{eq10}
\begin{aligned}
\varepsilon_{i+1}+\varepsilon_{i-1}&=(2-m\omega^2)\varepsilon_i\sec\theta,\\
\varepsilon_{i+1}-\varepsilon_{i-1}&=\omega\varepsilon_i\text{ cosec }\theta,
\end{aligned}
\end{equation}
\noindent
with $\varepsilon_0=\varepsilon_N-f=d$, to account for the boundary conditions of \eqref{eq7} in terms of $\varepsilon$. We try a solution of the form $\varepsilon_i=a^i\varepsilon_0$, which when substituted in \eqref{eq10} gives
\begin{equation}\label{eq10_0}
    a^{\pm1}=\left(1-\frac{m\omega^2}{2}\right)\sec\theta\pm\frac{\omega}{2}\text{cosec }\theta.
\end{equation}
Using the identity $aa^{-1}=1$ and further simplifications using equations \eqref{relation16}, we get from eq \eqref{eq10_0}
\begin{equation}\label{eq10_1}
    (\sec\theta-1)(2-f)=-2(d+1)\pm\sqrt{4(d+1)^2+f^2}.
\end{equation}
Again, eliminating $\omega$ from \eqref{relation16}, we get
\begin{equation}\label{eq10_2}
    mf^2(1+\cos\theta)=(d+1)(2d+f).
\end{equation}
Thus, the solution to equations \eqref{eq10_1} and \eqref{eq10_2} gives the bond configuration of the part of the chain with constant (or nearly constant) bond angle $\theta$ and bond length $\Delta$, having a `typical' radius $r=\frac{d+1}{2}\text{cosec}(\frac{\theta}{2})$ rotating with an angular velocity $\omega=\frac{f\sin\theta}{d+1}$.

In the overdamped limit, it has been shown that feasible solutions for $\Delta$ exist only within the range $0<f<2$. We studied the trajectories in details only for the parameter range $0<f<2$ to be consistent with the overdamped case. Although physically feasible solution does exist even for $f>2$ in inertial systems, simulations reveal that such trajectories are stable only for a very narrow range of parameter values, and those stable trajectories are qualitatively similar to those obtained for $0<f<2$. We show that, in this limit, equations \eqref{eq10_1} and \eqref{eq10_2} give only a pair of feasible solutions for $\theta$. 
Out of these two solutions (with different $d$, in general), for one the $\theta$ is acute, and for the other, $\theta$ is obtuse. To show this, we first observe that the LHS of \eqref{eq10_2} is non-negative and thus, from the non-negativity of the corresponding RHS, it follows that $d\geq-\frac{f}{2}$, with equality attained in the overdamped ($m=0$) case. Thus, for every possible choice of $\theta\in[0,\pi]$, we get a unique $d$ from eq\eqref{eq10_2}. Similarly, for any $d\geq-\frac{f}{2}$, eq\eqref{eq10_1} generates two solutions for $\theta$, one acute ($\sec\theta\geq1$) and the other obtuse ($\sec\theta\leq-1$), depending on the sign in front of the square root in RHS. Moreover, for any $\theta \in [0,\pi/2)$ eq~\eqref{eq10_2} provides a unique range of $d$ and for this range of $d$ eq~\eqref{eq10_1} only provides an acute solution. Similarly, for $\theta \in (\pi/2, \pi]$ eq~\eqref{eq10_1} only provides an obtuse solution for for the range of $d$ obtained from eq~\eqref{eq10_2}. Thus, these two operations together map any acute $\theta$ to another acute $\theta$ and any obtuse $\theta$ to another obtuse $\theta$. In this case, Brouwer's fixed-point theorem \cite{brouwer1911abbildung} guarantees at least one solution for each kind. In \hyperref[sec:sec5]{\blue{Supplementary S5}}, we have formally shown that both these solutions are unique.

We simulated this system for $N=100$ in the parameter range $0<f\leq2$ and $10^{-2}\leq m\leq 10^4$. Also, we restrict the simulations for $m > 0.01$, as the the chain trajectory and conformation was found to be quantitatively similar to the overdamped case for smaller $m$. In this range, we observed that for most cases, the chain conformation is such that $\theta$ is acute. The obtuse value of $\theta$ appears only for very small $m$ ($\lesssim0.04$), as the overdamped limit is approached. In such cases, the obtuse $\theta$ manifests in simulations as a distinct inner core within the more prominent outer ring formed by small $\theta$, as observed in purely overdamped systems~\cite{sadhu2025active}. Thus the simulations reveal that the 'core formation' is typically a feature of overdamped chains and the obtuse solution becomes unstable even for small inertia, where only the acute solution is stabilized. Since the core formation has already been studied in detail in \cite{sadhu2025active}, we focus only on the acute solution in this section.

We now derive an approximate solution for $d$ and $\theta$ to compare with simulations. To start with, we assume that $f\ll1$. For convenience, we introduce a variable $\xi = \frac{f}{2(d+1)}$, such that $\xi\le\frac{f}{2-f}\ll1$ (since, $d\geq-\frac{f}{2}$). Thus, we can simplify eq \eqref{eq10_1}, keeping only terms upto $\mathcal{O}(\xi^2)$ to obtain 
\begin{equation}\label{i}
    \sec\theta=1+\frac{f^2}{4(d+1)(2-f)}.
\end{equation}
Note that we have only considered the positive root (acute solution) in \eqref{eq10_1}. Plugging \eqref{i} in eq \eqref{eq10_2} and keeping terms only upto $\mathcal{O}(f^2)$, we get the following quadratic equation in $d$:
\begin{equation*}
    (d+1)^2-\left(1-\frac{f}{2}-\frac{f^2}{4(2-f)}\right)(d+1)-\left(m+\frac{1}{8}\right)f^2=0,
\end{equation*}
solving which we get the feasible solution as
\begin{equation}\label{ii}
    d'=-\frac{1}{2}\left\{\left(\mu+\frac{f}{2}\right)-\sqrt{\left(\mu-\frac{f}{2}\right)^2+4mf^2}\right\},
\end{equation}
where $\mu=1+\frac{f^2}{4(2-f)}$. $\Delta'=1+d'$ is a good approximation of $\Delta$ only for small $f$, but fails for large $f$. More specifically, we can see from equations \eqref{eq10_1} and \eqref{eq10_2} that as $f\to2$, $\theta\to\frac{\pi}{2}$ and $\Delta\to f\sqrt{\frac{m}{2}}$, whereas $\Delta'\to0$. Hence, $\Delta'$ fails to converge to the correct limit as $f\to2$. To fix this, we go back to the initial assumption of small $\xi$ and observe that when $m$ is large, $\Delta$ is large, and hence $\xi$ is small even for large $f$. Thus, the worst violation of eq \eqref{i} comes from the overdamped limit for large $f$. However, we note that in the overdamped limit, the acute solution for $\theta$ is of the form $\tan^{-1}(\frac{f}{2-f})$ \cite{sadhu2025active}. This observation motivates us to approximately express eq \eqref{i} in a similar form, as
\begin{equation}\label{iii}
    \theta \approx \tan^{-1}\left\{\frac{f}{\sqrt{2(d'+1)(2-f)}}\right\}
\end{equation}
under the small $\xi$ assumption, where $d'$ is given by eq \eqref{ii}. Now taking the limit $m\to0$, we get $d'\to-\frac{f}{2}$ and hence $\theta\to\tan^{-1}(\frac{f}{2-f})$ which happens to be the correct overdamped solution. In fact, the value of $\theta$ obtained from eq \eqref{iii} shows excellent match with simulated trajectories for different values of $m$ and $f$, as shown in Fig. \ref{fig4}(d). Using this value of $\theta$, we can also find the $\Delta$ from eq \eqref{eq10_2} as
\begin{equation}\label{iv}
    \Delta=d+1=\frac{1}{4}\left\{2-f+\sqrt{(2-f)^2+8mf^2(1+\cos\theta)}\right\}.
\end{equation}
One can check that this value of $\Delta$ actually converges to the correct limit as $f\to2$, unlike $\Delta'$ and hence is a good global approximation for the unperturbed bond length.
Fig \ref{fig4}(e) shows the comparison between the analytical values of $\Delta$ and those obtained from simulations for different values of $m$ and $f$. We observe that for small values of $m$, the bond length decreases with active force, whereas for large values of $m$, bond length typically increases with active force, with the crossover happening around $m\approx0.4$, where the dimensionless bond length roughly remains constant at $1$.

Going back to eq \eqref{eq10_0} and simplifying using relations \eqref{relation16}, we get $a^{\pm1}=\sqrt{1+\xi^2}\pm\xi$. Thus $a^{-1}\in(0,1)$, with $a^{-1}\to0$ as $\xi\to\infty$ (or, $f\to0$) and $a^{-1}\to1$ as $\xi\to0$ (or, $m\to\infty$), and 
consequently, $a>1$. This suggests that the deviations $\varepsilon_i$ grow towards the head of the chain $(i=N)$. Hence, this solution applies only to the head region of the chain, where the deviations obey $\varepsilon_{N-i}=a^{-i}\varepsilon_N$; $\varepsilon_N=d+f$, where $\theta$ remains constant.  One can see this relaxation behavior in simulations, in the head region of the chain as shown in Fig \ref{fig4}(f-g). However, in the tail section, $\theta$ does not remain constant any more, and so the analysis done so far, assuming a constant $\theta$, does not apply. This is precisely what is observed in simulations (see Fig \ref{fig4}(c)). To get the solution for the tail, we can no longer assume constancy of $\theta$, hence we follow the exact same procedure as done in \cite{sadhu2025active}, where we redo the analysis, assuming small deviations $\eta_i=\theta-\theta_i$ in the bond angles in addition to $\varepsilon_i$. Again from eq \eqref{eq7} we get

\begin{equation}\label{eq11}
\begin{aligned}
    (\varepsilon_{i+1}+\varepsilon_{i-1})\cot\theta+(m\omega^2-2)\varepsilon_i\text{ cosec }\theta&=(d+f)\eta_i+d\eta_{i-1},\\
    (\varepsilon_{i+1}-\varepsilon_{i-1})\tan\theta-\omega\varepsilon_i\sec\theta&=-(d+f)\eta_i+d\eta_{i-1},
\end{aligned}
\end{equation}
with $\varepsilon_0=\varepsilon_N-f=d$ as before and $\eta_0=\eta_{N-1}=0$, where we have retained only the first order terms in $\varepsilon_i$ and $\eta_i$. Rearranging, we get
\begin{equation}\label{eq12}
\begin{aligned}
    2(d+f)\eta_i&=A_-(\varepsilon_{i+1}-a\varepsilon_i)+A_+(\varepsilon_{i-1}-a^{-1}\varepsilon_i),\\
    2d\eta_{i-1}&=A_+(\varepsilon_{i+1}-a\varepsilon_i)+A_-(\varepsilon_{i-1}-a^{-1}\varepsilon_i),
\end{aligned}
\end{equation}
where $A_\pm=\cot\theta\pm\tan\theta$. Further eliminating $\eta_i$, we get
\begin{equation}\label{eq13}
\begin{aligned}
    &(d+f)A_+\varepsilon_{i+2}-\{a(d+f)A_++(a^{-1}(d+f)+d)A_-\}\varepsilon_{i+1}\\
    &+\{a^{-1}dA_++((a+1)d+f)A_-\}\varepsilon_i-dA_+\varepsilon_{i-1}=0,
\end{aligned}
\end{equation}
which has to be satisfied for all $i$. Trying the ansatz $\varepsilon_i \sim \alpha^i$ in eq \eqref{eq13},we get a cubic equation in $\alpha$. However, $\alpha=a$ is a trivial root corresponding to $\eta_i=0\;\forall i$. Factoring out this trivial root, we get the following quadratic equation after some simplification
\begin{equation}\label{eq14}
    \alpha^2-p\alpha+q=0,
\end{equation}
where $p=(a^{-1}+\frac{d}{d+f})\cos2\theta$ and $q=\frac{a^{-1}d}{d+f}$. The roots of this equation are $b_\pm=(p\pm\sqrt{D})/2$, where $D=p^2-4q$ is the discriminant. Thus, the tail solution is of the form $\varepsilon_i=u_+b_+^i+u_-b_-^i$, where $u_\pm$ are constants to be determined from boundary conditions. We do have the necessary boundary conditions at the tail end, but usually the bond angle deviations are too large at the tail for our boundary conditions (which were derived for small deviations) to hold. Hence, these constants are two free parameters for fitting the data from simulations.

We now examine whether the tail solution is physically realizable. 
Note that the fact that $d\ge-f/2$ implies
$|q|<1$. Also, $q=b_+b_-$ is the product of the two roots of \eqref{eq14}. Thus, at least one of the roots must be less than one in magnitude. Thus, in any case, we get at least one solution to eq \eqref{eq14}, for which the bond perturbations ($\varepsilon_i$, $\eta_i$) relaxes to zero towards the interior of the chain. It turns out that in fact both $|b_\pm|<1$ for the feasible range of parameters. Depending on the sign of the discriminant $D$, we get two different types of relaxation behavior:

(i) $D\ge0$: In this case, the roots $b_\pm$ are real. Thus the typical relaxation behavior is dictated by the dominant root. If the dominant root is positive, deviations $\varepsilon_i$ exponentially relax to zero keeping the same sign, while if it is negative, the deviations exponentially relax to zero alternating sign on successive bonds. Fig \ref{fig4}(f) shows the case for small $m=0.38$ and $f=0.4$, where $d<0$ (consequently, $q<0$). Thus the two roots have opposite signs. It turns out that in this case, the positive root $b_+$ dominates the relaxation behavior in the tail region as can be seen from the plot. Correspondingly, $\Delta_i$ monotonically relaxes to $\Delta$ similar to the head region as seen in Fig \ref{fig4}(c).

(ii) $D<0$: In this case the two roots are complex conjugates of each other. Hence the tail solution can be expressed as $\varepsilon_i=\zeta|b|^i\cos(i\phi+\psi)$, where $|b|=|b_\pm|$, $\phi=\arg(b_+)$ and $\zeta$, $\psi$ are constants to be determined which are related to $u_\pm$ as $\zeta=2|u_\pm|$ and $\phi=\arg(u_+)$. This is typically the case for large $m$, like the one shown in Fig \ref{fig4}(g) with $m=61.6$ and $f=0.9$, where the bond length deviations show oscillatory relaxation to zero with wavelength $2\pi/\phi$ and the amplitude decay exponentially as $|b|^i$.

The subtle difference in relaxation behavior is also reflected in the steady state configuration of the chain (Fig. \ref{fig4}(a-b)). Fig \ref{fig4}(a) corresponds to the case (i), where both the head and the tail regions slightly bend inward, whereas Fig \ref{fig4}(b) corresponds to the case (ii), where there is a distinctive `straightening' of the tail owing to the oscillatory relaxation. Also the inward curling of the head section is more pronounced because of the longer relaxation length (Fig \ref{fig4}(g)).

Finally, we close this section with a discussion on the chain length $N$, required to actually observe rigid solutions analyzed in this section. Our simulations indicate parameter regions where a rigid, circular solution is unstable. This is reflected as missing data points in Fig \ref{fig4}(d), for $N=100$. 
We have verified from simulations that increasing $N$ enlarges the parameter region of stable rigid steady states and decreasing $N$ shrinks this region.
The reason for this instability is purely structural, unlike the case for obtuse $\theta$.
Note that we derive different solutions of $\theta$ for the head and the tail ends, and the head solution relaxes towards the tail and vice versa. Thus, one condition for the coexistence of both solutions in a single chain is that the perturbations originating on each end should decay before reaching the other end. If the chain is shorter than the decay lengths, the perturbations from both ends can interact and destabilize the rigid steady state. 

The typical relaxation length of the chain is $\lambda=\frac{1}{\log a}-\frac{1}{\log b_m}$, where $b_m=\max(|b_+|,|b_-|)$. The first term corresponds to the head region and the second term to the tail region (the minus sign is because $0<b_m<1$). Thus, if $a$ and $b_m$ approaches $1$, $\lambda$ diverges and hence a longer chain is required sustain the  rigid solution. This is precisely what happens when the $(m,f)$ pair is large or small. Recalling that $a=\sqrt{1+\xi^2}+\xi$, where $\xi=\frac{f}{2\Delta}$; when $m$ is large, $\Delta$ is large and hence $\xi$ is small. Similarly, when $m$ is small, $\xi\approx\frac{f}{2-f}$ which is small for small $f$. Thus in these limits, $a\to1$ and hence the first term in $\lambda$ diverges. Additionally, for $f\to0$ or $2$, $|\cos2\theta|\to1$ and hence, $b_m\to a^{-1}$. Thus the second term in $\lambda$ also diverges. For the parameter values, where $\lambda\gg N$, the rigid solution is destabilized, as observed in simulations.

\section{Summary and Outlook}

In this work, we have studied the dynamics of active chains with a particular focus on the role of inertia. Our analysis has been carried out in the zero-noise limit. However, as demonstrated previously for overdamped chains~\cite{sadhu2025active}, the main findings are expected to remain valid in the presence of weak finite noise. We derived approximate analytical expressions for the bond lengths and bond angles in both the shortest chain ($N=3$) and the long-chain limit, and verified these predictions through numerical simulations. Furthermore, for short chains, we performed a linear stability analysis of the steady-state periodic solutions over a range of parameters and found excellent agreement with the numerical results. For long chains, the existence of multiple steady-state solutions gives rise to two distinct chain configurations characterized by different curvature distributions. In this regime as well, the analytical predictions for the chain shape show excellent agreement with simulations. Finally, we investigated the instability observed in long chains and identified its underlying structural origin.

Our study demonstrates that the intrinsic coupling between chain geometry and activity, implemented through the follower-force mechanism, gives rise to emergent periodic dynamics even in the simplest active-chain models. Such periodic behaviour is observed in a range of physical and biological systems, ranging from clamped, active filaments~\cite{chelakkot2014flagellar, de2017spontaneous, ling2018instability, Fatehiboroujeni2018, fily2020buckling, sangani2020elastohydrodynamical}  to motor-filament assays~\cite{shee2021semiflexible, yadav2024wave, ng2025active, collesano2022active}.
In particular, we find that the inclusion of inertia does not immediately suppress this periodic behaviour. Instead, the periodic states persist over a finite range of inertial parameters. Our analysis identifies the threshold beyond which inertia destabilizes these periodic motions, both for short and long active chains. 


Our study can be extended in several directions. One particularly interesting question is to investigate how the dynamics of the chain are modified in the presence of an external drive. In particular, it would be useful to understand how inertia influences the resulting dynamics. The model can be enriched by adding internal dynamical states and additional active interactions between them,  similar to swarmalator models~\cite{o2017oscillators, sar2025swarmalators, kreienkamp2025synchronization}. Additionally, our study in the presence of inertia will be relevant in the context of recent advancements in robotic systems inspired by studies in active matter~\cite{7hcf-p1yk, paramanick2024uncovering, son2025emergent, giardina2026robotectonics}.

\begin{acknowledgments}
\noindent
We thank Amitabha Nandi and Anirban Sain for helpful discussions. SS thanks Pampa Dey for helping towards the preparation of this manuscript and the supplementary information.
\end{acknowledgments}

\bibliographystyle{apsrev4-2}
\bibliography{Citation-Cluster}

\clearpage

\setcounter{page}{1}
\setcounter{figure}{0}
\setcounter{table}{0}
\setcounter{section}{0}
\setcounter{equation}{0}

\renewcommand{\thepage}{S\arabic{page}}
\renewcommand{\thefigure}{S\arabic{figure}}
\renewcommand{\thetable}{S\arabic{table}}
\renewcommand{\thesection}{S\Roman{section}}
\renewcommand{\theequation}{S\arabic{equation}}

\renewcommand{\thesection}{S\arabic{section}} 

\renewcommand{\thesubsection}{\thesection.\arabic{subsection}}

\centerline{\LARGE \bf Supplementary Information}
\vspace{0.5cm}

\section{Root bounds of Equation(9) from Main Text}\label{sec:sec1}
\noindent
Equation(9) from the main text reads
\begin{equation}\label{eq:eq1}
    F_{f} (\kappa) = (f-2)^2\kappa^3 + (f^2-6f+6)\kappa^2+f^2\kappa-(f^2-2f+2) = 0.
\end{equation}
We know that there can be only one value of $\kappa$ which is feasible, since $\kappa$ is the unique limiting value of $m\sin^{2}\theta$ when $m\to \infty$. This also means that $\kappa$ has to be non-negative. We have to find out the bounds of this feasible $\kappa$ and the condition(s) on $f$, if any, for feasibility.\\
\ 
The following equations (7), (8a), and (8b) from the main article will be required - 

\begin{equation}\label{eq:eq1_1}
m \omega^2=2 - \frac{\cos \theta(2-\cos \theta)}{m \sin ^2 \theta},
\end{equation}

\begin{equation}\label{eq:eq1_2}
        d_{+}=d_{1}+d_{2} = \frac{(f-2)(2-\cos\theta)}{\cos\theta(1-\frac{2-\cos\theta}{m\sin^2\theta})} - 2 ,
\end{equation}

\begin{equation}\label{eq:eq1_3}
        d_{-}=d_{1}-d_{2} = \frac{f}{1+\frac{2-\cos\theta}{m\sin^2\theta}}.
\end{equation}

\noindent
Using the underdamped limit ($m\to \infty$, $m\sin^2\theta\to \kappa$) in equations \eqref{eq:eq1_1}, \eqref{eq:eq1_2}, and \eqref{eq:eq1_3} we get, 
\begin{subequations}\label{eq:eq2}
    \begin{align}
        m\omega^2 &= 2-\frac{1}{\kappa}, \label{eq:eq2a_1}\\
        \Delta_{1} = d_{1} + 1 &= \frac{(f-1)\kappa^2 - \kappa}{\kappa^2-1},\label{eq:eq2b_2}\\
        \Delta_{2} = d_{2} + 1 &= \frac{(f-1)\kappa - \kappa^2}{\kappa^2-1}. \label{eq:eq2c_3}
    \end{align}
\end{subequations}
Feasibility requires $m\omega^2$ and the bond lengths ($\Delta_{1}, \Delta_{2}$) to be non-negative . Using the first condition in eq.\ref{eq:eq2a_1}, we get that $\kappa \ge 0.5$. Now, making use of the other two conditions in equations \eqref{eq:eq2b_2}, \eqref{eq:eq2c_3}, respectively, we get the following conditions on $\kappa$ for different ranges of $f$ as follows.\\
\ 
\begin{table}[h]
\centering
\begin{tabular}{|c|c|c|}
\hline
& Condition on $\kappa$ & Range of $f$ \\
\hline
(i)  & $0.5 < \kappa < 1$ & $0 < f \le 1.5$ \\
\hline
(ii) & $f - 1 \le \kappa < 1$ & $1.5 < f < 2$ \\
\hline
(iii) & $1 < \kappa \le f - 1$ & $f > 2$ \\
\hline
\end{tabular}
\caption{Feasibility Conditions}
\label{tab:Table1}
\end{table}

\noindent
These are just the feasibility conditions; we still have to check whether there exist a solution to equation \eqref{eq:eq1} under these conditions. First of all, we note that there $F_{f}(0.5) = -\frac{f^2}{8}<0$ and $F_{f}(1) = 2 (f-2)^2>0$ $\forall$ $f\neq 0,2$, which means there exists at least one root in the interval $\kappa \in (0.5, 1)$. Further, we also note that $F_{f}^{'}(\kappa) = 3(f-2)^{2} \kappa^2 + 2(f^2-6f+6)\kappa + f^{2} = f^{2}(1-\kappa) + 3\kappa(\kappa+1)(f-2)^{2}$ is strictly positive in the interval $\kappa\in(0.5, 1)$, which means that  $F_{f}(\kappa)$ is strictly monotonic. Thus, there exists exactly one root in this interval. Moreover, $F_{f}(f-1) = f^2(f-2)^3<0$ for $f<2$. Hence, $\kappa\in(f-1,1)$ provides a stricter bound on $\kappa$ for $1.5<f<2$. In fact, a tighter lower bound is given by $\kappa\in(\frac{1}{3-f},1)$ for $f\in(1,2)$.\\
\ 

\noindent
This shows that there exists a root of eq.\eqref{eq:eq1} which satisfies conditions (i) and (ii) (Table \ref{tab:Table1}). However, no solution of \eqref{eq:eq1} satisfies condition (iii), because $F_{f}$ can be written as $F_{f}(\kappa) = (f-2)^2(\kappa-1)(\kappa + 1)^2 + 2[(f-1)(\kappa - 1)+(f-2)]((f-1)-\kappa)$, which is strictly positive under condition (iii), and hence there is no root. To sum up, we find that the feasible solution of eq.\eqref{eq:eq1} exists only for $f\in(0,2)$, and it lies in the interval $\kappa\in(\max(0.5,\frac{1}{3-f}), 1)$ with $\kappa=0.5$ for $f=0$ and $\kappa=1$ for $f=2$.

\section{Approximate solution of Equation(9) from Main Text}\label{sec:sec2}
\noindent
Equation \eqref{eq:eq1} is a cubic equation and hence it can be solved exactly in terms of radicals using Cardano's method, but such an expression would be extremely cumbersome and highly impractical, especially since eq.\eqref{eq:eq1} has multiple real roots for a certain feasible range of $f$. Hence, we look for closed-form approximate solutions which are cleaner and practically more useful to estimate $\kappa$.\\
\ 

\noindent
Since we know the bounds of $\kappa$, any interval based numerical method can give the required solution upto any arbitrary precision. However, analytical approximations can only take us `sufficiently close' to the exact value, if the global error is considered. There is no systematic recipe, as we know of, to construct a good global approximant (analogous to the `local' Taylor series approximation), which can make the global error arbitrarily small. Hence, here we present a few methods that seems to work well globally based on trial and error.

\subsection{Pad\'e approximation}\label{sec:sec2a}
\noindent
When it comes to global approximation of a function over an interval, Pad\'e approximants do a better job than Taylor series in minimizing the global error. For our problem, we can find out the Taylor approximation at the end points of the interval in $f$, that is $f\in[0,2]$ up to any order. Using the coefficients of such Taylor approximants, we can generate Pad\'e approximation of any order of the function in that interval. To get the local Taylor series at the end points, we make use of the implicit equation \eqref{eq:eq1} using the following general procedure.\\
\ 

\noindent
Suppose, we want to generate the local Taylor series approximation of the root $\kappa (f)$ of eq.\eqref{eq:eq1} around some $f=f_{0}$. Further, suppose $\kappa(f_{0}) = \kappa_{0}$ is known. Then, we can iteratively plug in the series ansatz $\kappa^{(n)}(f) = \kappa^{(n-1)}(f) + a_n(f-f_0)^n$, $n\ge1$ with $\kappa^{(0)}(f) = \kappa_0$ to $F_f(\kappa)$ starting from $n=1$ and solve for $a_n$ by setting the coefficient of the most dominant term of $F_{f}(\kappa)$, expressed in powers of $(f-f_{0})$ (i.e. the lowest degree term in $(f-f_0)$) to zero. Thus, we generate the Taylor coefficients, one at a time, such that $F_{f}(\kappa^{(n)}(f))$ is at least  $\mathcal{O}((f-f_0)^{n+1})$. Following this procedure, we get the Taylor series approximation of $\kappa(f)$ at $f=0$ and $f=2$ as follows 

\begin{subequations}\label{eq:eq3}
    \begin{align}
        \kappa(0) &= \frac{1}{2} + \frac{f^2}{72} + \frac{f^3}{72} + \frac{73}{7776}f^4 + \mathcal{O}(f^5),\label{eq:eq3a_1}\\
        \kappa(2) &= 1+(f-2) + (f-2)^2 + (f-2)^3 + \frac{5}{4}(f-2)^4+\mathcal{O}((f-2)^5). \label{eq:eq3b_2}
    \end{align}
\end{subequations}
A Pad\'e approximation $\kappa_{[m,n]}(f)$ of $\kappa(f)$, is a rational approximation of the form $\mathcal{P}_m(f)/\mathcal{Q}_n(f)$, where $\mathcal{P}_m$ and $\mathcal{Q}_n$ are polynomials in $f$ of degrees $m$ and $n$ respectively, and $\mathcal{Q}_n(0)=1$ without loss of generality. Thus, there are total $m+n-1$ coefficients, which are to be obtained. In this case, since we have the taylor series at both the ends, we can equate the first $p$ and $q$ coefficients of $\kappa_{[m,n]}(f)$, expanded around $f=0$ and $f=2$ respectively to the corresponding coefficients from eq.\eqref{eq:eq3} to obtain a total of $p+q$ linear equation. If $p+q=m+n-1$, this system becomes closed and guarantees a unique solution (if full rank). Using this method, we obtained a few lower-order, near-diagonal Pad\'e approximants out of which the following works the best \\
\ 
\begin{equation}\label{pade}
    \kappa^{pade}_{[3,4]}(f) = \frac{8(f^3-8f^2 + 18f - 18)}{f^4+16f^3-120f^2+288f-288},
\end{equation}
\noindent
with the maximum relative error less than $10^{-3}$.\\
\ 
As a consistency check, we must ensure that the proposed Pad\'e approximant has no `spurious' pole in the region of interest, i.e. the denominator of \eqref{pade} (denoted by $\mathcal{Q}_4(f)$) has no root in $f\in(0,2)$. It is indeed the case here, because $\mathcal{Q}_4(0)=-288$, $\mathcal{Q}_4(2)=-48$ which are both of same sign and $\mathcal{Q}^{'}_4(0)=4(2-f)^2(16+f)+32$ is strictly positive for $f\in(0,2)$. Hence the formula \eqref{pade} is perfectly well behaved in the desired range.\\
\
\noindent
For this particular problem it is also possible to get a multi-point Pad\'e approximation by comparing a Taylor series at more than two points. It is so, because the original equation \eqref{eq:eq1} is quadratic in $f$, and hence, $f(\kappa)$, which is the inverse function of $\kappa(f)$, can be exactly obtained. Hence, we can obtain as many $(f,\kappa)$ solution pairs for eq.\eqref{eq:eq1} as we want. From there, we can obtain the Taylor approximations at those points using the procedure described above. However, we have not mentioned such approximants here for the sake of simplicity.

\subsection{Quadratic approximation of eq.\eqref{eq:eq1}}\label{sec:sec2b}
Suppose we want to find an approximate solution to eq.\eqref{eq:eq1} around $f=f_{0}$, where $\kappa(f_{0}) = \kappa_{0}$. Then, we can expand $F_{f}(\kappa)$ in terms of the form $(f-f_{0})^{\alpha} (\kappa-\kappa_{0})^{\beta}$. The idea, then, is to generate an approximation to $F_{f}(\kappa)$ by neglecting (or modifying) higher-order terms in $\alpha + \beta$ and setting the resulting expression to zero. The new equation thus obtained is tailored in such a way that it is easier to solve, and the solution obtained is expected to be an approximation to the true solution around $f=f_{0}$. We found that such approximations are better than the local Taylor series expansion for this particular problem. \\
\ 

\noindent
We applied this procedure to get an approximation of $\kappa$ around $f=0$ by expressing $F_{f}(\kappa)$ in terms of $f^{\alpha}(\kappa-0.5)^{\beta}$. Since, $\max(\alpha) = 2$ and $\max(\beta) = 3$, the least dominant term is $f^{2}(\kappa-0.5)^{3}$. A modification to the coefficient of this term leads to a trivial factorization. More specifically, $F_{f}(\kappa) - \frac{16}{27}f^{2}(\kappa - \frac{1}{2})^3$ has $(\kappa + 1)$ as one of it's factors. Hence, we can factor out this trivial root and the resulting quadratic in $\kappa$ has the solution,

\begin{equation}\label{quad2}
    \kappa_{2}^{quad}(f) = \frac{-27+27f-20f^2+3\sqrt{3(243 - 486f + 405f^2 -162f^3 + 25f^4)}}{108-108f+11f^2}.
\end{equation}
\noindent
It is to be noted that out of the two possible solutions, this the one which leads to the correct $\kappa$. This solution happens to be a very good approximation to $\kappa(f)$ in the range $f\in(0,1)$, but it fails beyond $f=1$. Another possible approach is to neglect the entire $(\kappa - \frac{1}{2})^3$ term and solve the remaining quadratic equation in $(\kappa - \frac{1}{2})$ to get the following root,

\begin{equation}\label{quad1}
    \kappa_{1}^{quad}(f) = \frac{(12 - 12f -f^2) - (f-2)\sqrt{3(108 - 108f + 47f^2)}}{96-96f+20f^2}.
\end{equation}

\noindent
Although there is a slight abuse of rigor here, because we are not neglecting the terms order by order, yet this approach leads to an excellent global approximation of $\kappa$ with a percentage error of less than 0.3\%. Moreover, it satisfies the end point conditions $\kappa_{1}^{quad}(0) = \frac{1}{2}$ and $\kappa_{1}^{quad}(2) = 1$ exactly.\\
\ 

\begin{figure}[htp]
    \centering

\includegraphics[width=0.7\linewidth]{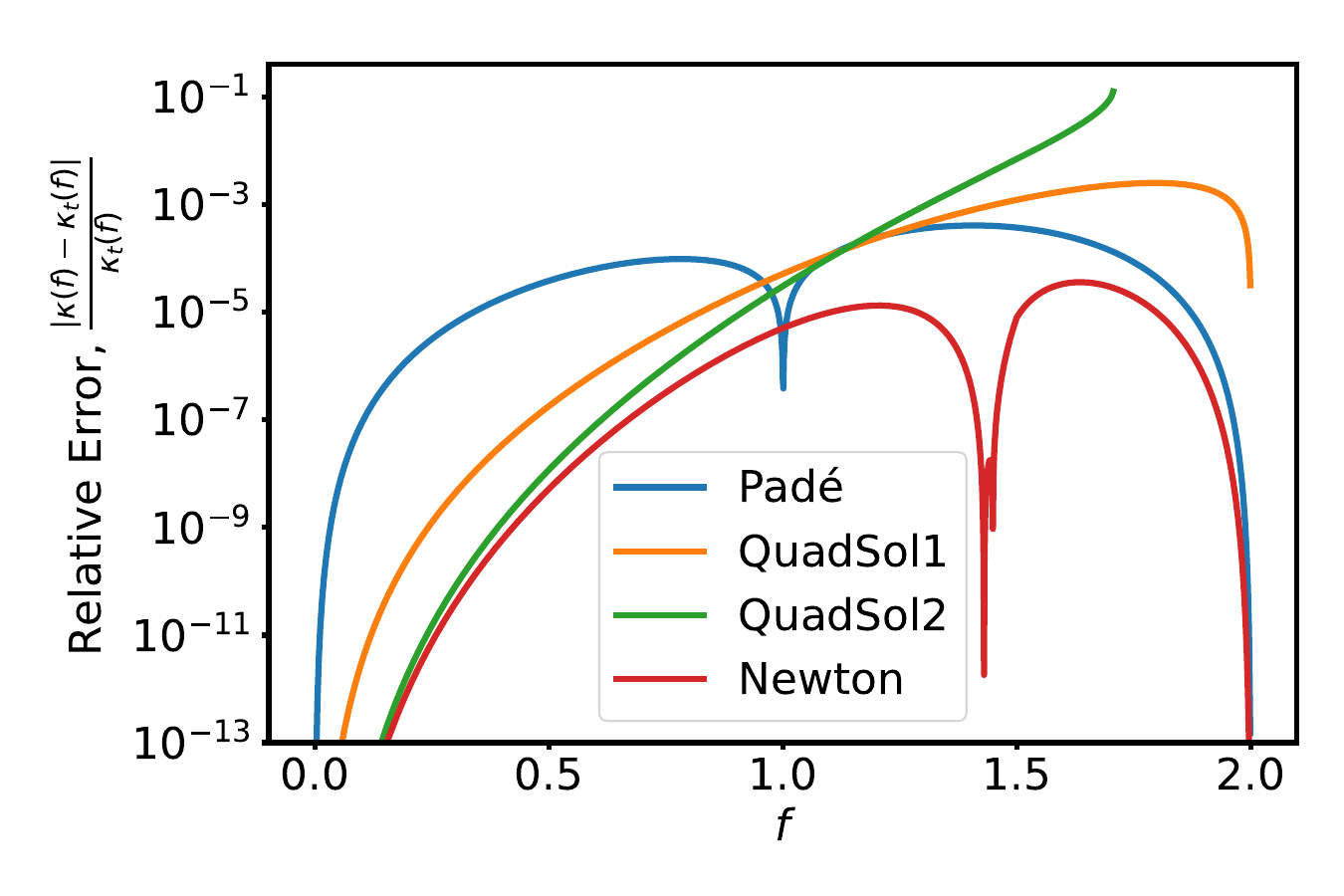}
    \caption{Comparison of relative error ($|\kappa-\kappa_t|/\kappa_t$) for different schemes over $f\in(0,2)$. $\kappa$ is the approximate value and $\kappa_t$ is the true value. Pad\'e refers to the $\kappa^{pade}_{[3,4]}$ solution as in eq.\eqref{pade}; QuadSol1,2 refers to the quadratic solutions $ \kappa_{1,2}^{quad}$ as in eq.\eqref{quad1} and \eqref{quad2} respectively; and Newton refers to two steps of Newton-Raphson starting from $\max(f-1,0.5)$.}
    \label{fig:figs1}
\end{figure}

\noindent
All these formulas discussed above are compared in figure \ref{fig:figs1} and it is seen that the Pad\'e is the best overall approximant. However, this the best we can do with analytical approximations. Just for comparison, only two iterations of the Newton-Raphson method starting from $\max(f-1,0.5)$ already produces lower relative error than any of these methods.

\section{Maclaurin series and Asymptotic expansion of $\cos\theta$ and $\omega^2$ in terms of $m$}\label{sec:sec3}
\noindent
Starting from equation(5) of the main article
\begin{equation}\label{eq10_1}
        d_1(d_1+1)+(d_2+f)(d_2+1)=0,
\end{equation}
\noindent
we plug in the values of $d_{1}, d_{2}$ from equations \eqref{eq:eq1_2}, \eqref{eq:eq1_3}, and after some simplification, we obtain the following sextic equation in $\cos\theta$
\begin{equation}
\begin{aligned}
&m^{2}(1 - F - 2m)\cos^{6}\theta+ 2m(1 + F - 3m)\cos^{5}\theta + ((1 - F) - 2(5 + 2F)m + 2(3 + F)m^{2} + 6m^{3})\cos^{4}\theta - \\&2(3 - 2F - (7 - F)m - 6m^{2})\cos^{3}\theta + (4(3 - F) + 2(1 + 2F)m - (15 + F)m^{2} - 6m^{3})\cos^{2}\theta - 2(2 + m)\\&(2 + 3m)\cos\theta + 2m(2 + m)^{2} = 0,
\end{aligned}
\end{equation}
where $F = \frac{f^2}{(2-f)^2}$. It turns out that $F$, which is the square of the overdamped angular velocity is the more natural variable rather than $f$ for the subsequent analyses.\\

\noindent
To get the series expansions, we can use the exact same procedure as described in section \ref{sec:sec2a}. Our series ansatz for Maclaurian series is of the form $\cos\theta = \sum_{l} a_{l}m^{l}$ and for asymptotic series is of the form $\cos\theta = \sum_{l}b_{l}m^{-l}$ where $l\ge0$. Following are the first few terms of the series constructed using this procedure - 

\begin{subequations}\label{cos}
\begin{align}
    &\cos\theta
    = m
    + \frac{1}{2}(1-F)m^{2}
    - \frac{1}{2}(1-F^{2})m^{3}
    - \frac{1}{8}\left(11 - 13F + F^{2} + 5F^{3}\right)m^{4}
    + \mathcal{O}(m^{5}),\label{cosl}\\
    &\text{when $m\to0$, and}\nonumber\\\nonumber\\
    &\cos\theta
    = 1 - \frac{\kappa}{2m}
    + \frac{(11+3F)(1+F) - 2(7+14F+3F^{2})\kappa - (25-2F-3F^{2})\kappa^{2}}
    {32\left(F + (3-F)\kappa + 3\kappa^{2}\right)m^{2}}
    + \mathcal{O}(m^{-3}),\label{cosr}
\end{align}
\end{subequations}
when $m\to\infty$. Now, plugging \eqref{cos} in the expressions \eqref{eq:eq1_1}-\eqref{eq:eq1_3}, we can get the corresponding series expansions. In particular, from \eqref{eq:eq1_1}, we get the Maclaurin and Asymptotic expansions of $\omega^2$ as follows -
\begin{subequations}\label{ww}
\begin{align}
    &\omega^{2}
    = F
    - F(1+F)\,m
    - \frac{F}{4}\left(3 - 6F - 5F^{2}\right)m^{2}
    + \frac{F}{4}\left(3 + 9F - 9F^{2} - 7F^{3}\right)m^{3}
    + \mathcal{O}(m^{4}),\label{wwl}\\
    &\text{when $m\to0$, and}\nonumber\\\nonumber\\
    &\omega^{2}
    = \left(2 - \frac{1}{\kappa}\right)m^{-1}
    + \frac{1 + (F-1)\kappa - (F+2)\kappa^{2}}
    {3(F+1) - 4F\kappa + (F-3)\kappa^{2}}\, m^{-2}
    + \mathcal{O}(m^{-3}),\label{wwr}
\end{align}
\end{subequations}
when, $m\to\infty$. We will require the expansions \eqref{ww} for constructing the Pad\'e approximation of $\omega^2$ in the main article.


\section{Linear Stability analysis for $N=3$}\label{sec:sec4}

Any rigid state of our active chain can be considered as a fixed point in the phase space formed by the bond lengths and angles and their derivatives. Note that the derivatives are present because there are acceleration terms due to inertia. This fixed point solution is linearly stable iff the corresponding Jacobian matrix is Hurwitz (all eigenvalues have negative real part). For the 3-bead system, we derived the following reduced Jacobian by setting the bond extension velocities to zero and bond angular velocities to $\omega$ -

\begin{equation}\label{jacobian}
    J =
    \begin{bmatrix}
        0&0&0&1&0&0&0 \\
        0&0&0&0&1&0&0 \\
        0&0&0&0&0&-1&1\\
        \frac{m\omega^2-2}{m}&\frac{\cos\theta}{m}&-\frac{(d_1+1)\omega}{m}&-\frac{1}{m}&0&2(d_1+1)\omega&0\\
        \frac{\cos\theta}{m}&\frac{m\omega^2-2}{m}&\frac{(d_2+1)\omega}{m}&0&-\frac{1}{m}&0&2(d_2+1)\omega\\
        -\frac{\omega}{(d_1+1)m}&\frac{\sin\theta}{(d_1+1)m}&\frac{(d_2+f)\cos\theta}{(d_1+1)m}&-\frac{2\omega}{d_1+1}&0&-\frac{1}{m}&0\\
        -\frac{\sin\theta}{(d_2+1)m}&-\frac{\omega}{(d_2+1)m}&-\frac{d_1\cos\theta}{(d_2+1)m}&0&-\frac{2\omega}{d_2+1}&0&-\frac{1}{m}
    \end{bmatrix},
\end{equation}

which works for both the 1D linear and 2D circular rigid steady states. Now, we focus on these two cases one by one. The analyses will be restricted to the feasible range $0<f<2$ only.

\subsection{Linear Case}\label{sec:sec4a}
\noindent
For the 1D linear case, we have $\omega=0$ and depending on the bond angle, we can have the following two steady states, namely the `extended' state with $\{d_{1} =  - f / 3,\ d_{2} =  - 2f / 3,\ \theta = 0;\ 0<f<1.5\}$ and the `folded' state with $\{ d_{1} =  - f,\ d_{2} = 0,\ \theta = \pi;\ 0<f<1 \}$. Substituting these values in \eqref{jacobian}, we get the characteristic equations for the eigenvalue $\lambda$ to be -
\begin{equation*}
    (1+m\lambda)(1+\lambda+m\lambda^2)(3+\lambda+m\lambda^2)(f^2-c_\pm\lambda-mc_\pm\lambda^2)=0,
\end{equation*}
with $c_+=9-9f+2f^2$ and $c_-=1-f$, where the signs `+' and `-' corresponds to the extended and the folded configurations, respectively. Clearly, stability requires $c_\pm<0$, which is impossible (see Section III.A.3 in \cite{sadhu2025active}), hence the linear states are unstable in the feasible parameter range. With the linear states out of the way, we now look towards the circular state.

\subsection{Circular Case}\label{sec:sec4b}
\noindent
Equation \eqref{jacobian} gives the most general form of the Jacobian for circular steady states, which cannot be simplified any further. This form of the Jacobian does not exhibit any trivial factorization of the corresponding eigenvalue equation, and hence needs to be treated with full generality. One of many methods, used to check whether a real full-rank matrix is Hurwitz is the Routh-Hurwith Criteria(\cite{routh},\cite{hurwitz}), which we briefly describe here as follows before applying to our problem:\\

Suppose, $\sum_{k=0}^na_k\lambda^k$ is the characteristic polynomial of a real-valued matrix $A_{n\times n}$, whose zeros are the eigenvalues of $A$. Then, the Necessary Condition for $A$ to be Hurwitz is that all the coefficients $a_k$ must be of same sign, and the Sufficient Condition is that all entries in the first column of the so called Routh Matrix (or, Routh Table) $R$ must be of the same sign. $A$ is Hurwitz iff it satisfies both the Necessary and the Sufficient condition. Moreover, the number of eigenvalues of $A$ with positive real part is exactly equal to the number of sign changes along the first column of $R$ while traversing from top to bottom, in case $A$ is not Hurwitz. The matrix $R_{(n+1)\times(\lfloor n/2\rfloor+1)}$ is constructed as follows: The coefficients $a_k$ are alternately placed in the first two rows of $R$ such that $R_{0,l}=a_{n-2l}$ and $R_{1,l}=a_{n-2l-1}$ for $l\in\{0,\dots,\lfloor\frac{n}{2}\rfloor\}$. For the remaining rows, $R_{k,l}=\frac{R_{k-1,0}R_{k-2,l+1}-R_{k-2,0}R_{k-1,l+1}}{R_{k-1,0}}$ for $k\in\{2,\dots,n\}$ and $l\in\{0,\dots,\lfloor\frac{n}{2}\rfloor\}$. During this construction, we take $a_k=0$ for any $k\notin\{0,\dots,n\}$. Also, during the construction of $R$, the rows can be scaled by any positive number, since that keeps the sign unaffected. By this particular construction (called the Routh construction) the Hurwitz condition simply becomes $\operatorname{sign}(a_k)=\operatorname{sign}(R_{k,0})=\operatorname{sign}(a_0)\ \forall k\in\{0,\dots,n\}$, which is very simple to check for our problem.\\

From simulations, we know that the circular state starts stable in the overdamped limit and becomes unstable at some critical $m^*$ as the mass is increased. We are mostly interested in the region where $m$ is large and $f$ is small, because in this parameter regime we expect some kind of power law dependence of $m^*$ on $f$, which we aim to obtain following the Routh procedure. We use the equations \eqref{eq:eq2} and \eqref{eq:eq3a_1} in \eqref{jacobian} and keep only the leading order term for each entry of the Jacobian. The leading order term here, is the term with the highest power in $m$ and lowest in $f$. It turns out that keeping only the leading term is sufficient for this analysis. The resulting Jacobian and the corresponding Routh Matrix are given as follows:

\begin{equation*}
\resizebox{\textwidth}{!}{%
    $J' =
    \begin{bmatrix}
        0&0&0&1&0&0&0 \\
        0&0&0&0&1&0&0 \\
        0&0&0&0&0&-1&1\\
        \frac{-2}{m}&\frac{1}{m}&-\frac{f}{3m\sqrt{2m}}&-\frac{1}{m}&0&\frac{2f}{3\sqrt{2m}}&0\\
        \frac{1}{m}&\frac{-2}{m}&\frac{f}{3m\sqrt{2m}}&0&-\frac{1}{m}&0&\frac{2f}{3\sqrt{2m}}\\
        -\frac{f}{3m\sqrt{2m}}&\frac{1}{m\sqrt{2m}}&\frac{f}{3m}&-\frac{2f}{3\sqrt{2m}}&0&-\frac{1}{m}&0\\
        -\frac{1}{m\sqrt{2m}}&-\frac{f}{3m\sqrt{2m}}&\frac{f}{3m}&0&-\frac{2f}{3\sqrt{2m}}&0&-\frac{1}{m}
    \end{bmatrix},
    \quad
    R =
    \begin{bmatrix}
        1&\frac{4}{m}&\frac{3}{m^2}&\frac{3}{m^4}\\
        1&\frac{3}{m}&\frac{3}{2m^2}&\frac{f^2}{6m^3}\\
        1&\frac{3}{2m}&\frac{3}{m^3}-\frac{f^2}{6m^2}&0\\
        1&\frac{1}{m}&\frac{f}{9m^2}&0\\
        1&\frac{6}{m^2}-\frac{5f^2}{9m}&0&0\\
        1&\frac{f^2}{9m}&0&0\\
        1-\frac{mf^2}{9}&0&0&0\\
        1&0&0&0
    \end{bmatrix}$.
}
\end{equation*}
Just like $J'$, the entries of $R$ are also kept only upto the leading order. Also, the rows are rescaled during construction, so that the first entry of every row, except the $7$th row are all one. Thus, the stability of this entire system boils down to the sign of the element $R_{6,0}=1-\frac{mf^2}{9}$. The system is stable if $R_{6,0}>0$, or $m<m^*=9/f^2$ and unstable otherwise, with the critical mass given by $m^*=9/f^2$. Moreover it also tells us that when the system becomes unstable, there are two sign changes along the first column of $R$, implying that there are two eigenvalues which crosses the imaginary axis to the right half plane. It turns out from simulation that this is a complex conjugate pair of eigenvalues.\\

Similarly, it can be shown using the same procedure, that the Jacobian \eqref{jacobian} is always Hurwitz in the overdamped limit, using the small $m$ expansions derived earlier as long as the taylor coefficients remain sublinear in $m$. This is an expected result, since we already know that the perfectly overdamped system is always stable \cite{sadhu2025active}, hence the full calculation is not shown here in details.

\section{Number of feasible solutions for large $N$}\label{sec:sec5}

The solution $x\equiv\cos\theta$ for the steady unperturbed part of the chain for large $N$ can be expressed as the solution of the fixed point equations $x=h_\pm(x)$, where $h_\pm=X\circ Y_\pm\circ\Delta$ is a composition of the following three implicit functions (of $z$), defined as follows:
\begin{subequations}\label{eq5.1}
    \begin{align}
        &\Delta(2\Delta-2+f)=mf^2(1+z),\label{eq5.1.1}\\
        &Y^2+4zY=f^2,\label{eq5.1.2}\\
        &\frac{1}{X}=1+\frac{z}{2-f}.\label{eq5.1.3}
    \end{align}
\end{subequations}
Note that $Y_\pm$ are the two solutions of eq \eqref{eq5.1.2} given by
\begin{equation}\label{eq5.2}
    Y_\pm(z)=-2z\pm\sqrt{(2z)^2+f^2}.
\end{equation}
All we have done is put equations (17) and (18) from main article in a more convenient form. We have already shown that for the parameter range $f\in(0,2)$, $x\in(-1,1)\setminus\{0\}$ and $h_+:[0,1]\to(0,1)$ corresponds to the acute solution and $h_-:[-1,0]\to(-1,0)$ corresponds to the obtuse solution. Also, Brouwer's fixed point theorem \cite{bfpt} ensures the existence of a solution in both cases. Here, we show that both these solutions are in fact unique.

We calculate the first two derivatives of the functions \eqref{eq5.1} as follows
\begin{subequations}\label{eq5.3}
    \begin{alignat}{2}
        &\Delta'=\frac{mf^2}{4\Delta-2+f},\qquad&&\Delta''=-\frac{4(\Delta')^2}{4\Delta-2+f};\label{eq5.3.1}\\
        &Y'=-\frac{2Y}{Y+2z},&&Y''=-\frac{Y'(Y'+4)}{Y+2z};\label{eq5.3.2}\\
        &X'=-\frac{X^2}{2-f},&&X''=-\frac{2XX'}{2-f}.\label{eq5.3.3}
    \end{alignat}
\end{subequations}
Now, we would like to determine the sign of each of these quantities. Note that by replacing the argument $z$ by $x$ in eq \eqref{eq5.1.1}, $\Delta(x)$ has the interpretation of bond length which has already been shown to hold the following
\begin{equation}\label{eq5.3_1}
    \Delta(x)=1+d\ge1-\frac{f}{2}>0.
\end{equation}
Thus, $4\Delta(x)-2+f\ge2-f>0$, and from eq \eqref{eq5.3.1}, we get
\begin{equation}\label{eq5.4}
    \Delta'(x)>0\quad\text{and}\quad\Delta''(x)<0.
\end{equation}
Note that these inequalities are strict only when $m>0$, which is exactly what we care for, because the case of $m=0$ has already been explicitly dealt with in \cite{sadhu2025active}, where we have shown that the two solutions are unique. $\Delta(x)$ is strictly concave and increasing in the feasible domain. Such functions are called concavity-preserving, because when operated on another concave function, the result of the composition is also concave. Further, from eq \eqref{eq5.2}, it follows that
\begin{equation}\label{eq5.5}
    Y_+(\Delta)>0\quad\text{and}\quad Y_-(\Delta)<-4\Delta,
\end{equation}
where $\Delta\equiv\Delta(x)$ is a shorthand introduced for notational brevity. Using relations \eqref{eq5.3_1} and \eqref{eq5.5} in \eqref{eq5.3.2}, we get
\begin{equation}\label{eq5.6}
    Y'_\pm(\Delta)<0,\quad Y''_+(\Delta)>0\quad\text{and}\quad Y''_-(\Delta)<0.
\end{equation}
Similarly, from \eqref{eq5.1.3} and \eqref{eq5.3.3}, we get
\begin{equation}\label{eq5.7}
    X'(Y_\pm)<0,
\end{equation}
where $Y_\pm\equiv Y_\pm(\Delta(x))$
Now we prove the uniqueness of the two fixed points of $h_\pm(x)$ separately.
\subsection{Uniqueness of the Acute solution}
$h'_+(x)=X'(Y_+)Y_+'(\Delta)\Delta'(x)>0$ by relations \eqref{eq5.4}, \eqref{eq5.6} and \eqref{eq5.7}. Again, using relations \eqref{eq5.1}, \eqref{eq5.3}, \eqref{eq5.3_1} and \eqref{eq5.5}, we get the following relations
\begin{subequations}
    \begin{align*}
        -\frac{1}{X'(Y_+)}&=(2-f)\left(1+\frac{Y_+}{2-f}\right)^2>2Y_+\;,\\
        -\frac{1}{Y'_+(\Delta)}&=\frac{1}{2}+\frac{\Delta}{Y_+}>\frac{\Delta}{Y_+}\;,\\
        \frac{1}{\Delta'(x)}&=\frac{1+x}{\Delta}\left(2+\frac{2-f}{2\Delta-2+f}\right)>\frac{2}{\Delta}\;.
    \end{align*}
\end{subequations}
Multiplying and taking the inverse, we get $h_+'(x)<\frac{1}{4}<1$. What these results imply is that $y=h_+(x)$ is a strictly increasing function in the interval $(0,1)$ whose slope never exceeds $1$. Thus, once this function crosses the line $y=x$, it cannot cross back again, which means that the corresponding fixed point is unique. This argument can be formalized by applying Intermediate Value Theorem on the monotonic function $h_+(x)-x$.
\subsection{Uniqueness of the Obtuse Solution}
Let us define $g=X\circ Y_-$. Clearly, $g'(\Delta)=X'(Y_-)Y'_-(\Delta)>0$ by \eqref{eq5.6} and \eqref{eq5.7}. Taking derivative again, and using chain rule, $g''(\Delta)=X''(Y_-)(Y'_-(\Delta))^2+X'(Y_-)Y''_-(\Delta)$, which can be simplified using equations \eqref{eq5.3} to obtain
\begin{equation*}
    g''(\Delta)=-\frac{X^3}{2-f}\left(Y''_-+\frac{(Y')^3}{2(2-f)}\right)<0
\end{equation*}
where the last inequality results from the relation \eqref{eq5.6} and the fact that $X(Y_-)=\frac{2-f}{2-f+Y_-}<0$. Thus, like $\Delta(x)$, $g(\Delta)$ is also a concavity-preserving function in the feasible parameter space. Thus their composition $h_-(x)=g(\Delta(x))$ is also strictly concave and increasing ($h'_-(x)>0$ and $h''_-(x)<0$). Once the function $h_-(x)$ crosses the line $y=x$, it cannot turn back to cross it again. Hence, the solution to the fixed point equation $h_-(x)=x$ is unique.

\end{document}